\documentstyle[epsfig]{elsart}
\begin{document}
\begin{frontmatter}
\title{The STACEE-32 Ground Based Gamma-ray Detector}
\author{D.S. Hanna,$^1$} 
\author{D. Bhattacharya,$^2$}
\author{L.M. Boone,$^3$}
\author{M.C. Chantell,$^4$}
\author{Z. Conner,$^4$}
\author{C.E. Covault,$^{4,7}$}
\author{M. Dragovan,$^{4,8}$}
\author{P. Fortin,$^1$}
\author{D.T. Gregorich,$^5$}
\author{J.A. Hinton,$^4$}
\author{R. Mukherjee,$^6$}
\author{R.A. Ong,$^{4,9}$}
\author{S. Oser,$^{4,10}$}
\author{K. Ragan,$^1$}
\author{R.A. Scalzo,$^4$}
\author{D.R. Schuette,$^{4,11}$}
\author{C.G. Th\'{e}oret,$^{1,12}$}
\author{T.O. T\"{u}mer,$^2$}
\author{D.A. Williams,$^3$}
\author{J.A. Zweerink,$^{2,9}$}
\address{$^1$ Department of Physics, McGill University, Montreal, QC H3A2T8, Canada}
\address{$^2$ Institute of Geophysics and Planetary Physics, 
University of California, Riverside, CA 92521, USA}
\address{$^3$ Santa Cruz Institute for Particle Physics,
University of California, Santa Cruz, CA 95064, USA}
\address{$^4$ Enrico Fermi Institute, University of Chicago, Chicago, IL 60637, USA}
\address{$^5$ Department of Physics and Astronomy, California State University, LA,
Los Angeles, CA 90032, USA}
\address{$^6$ Department of Physics \& Astronomy,
Barnard College and Columbia University, New York, NY 10027, USA}
\address{$^7$ Present Address: Department of Physics, 
Case Western Reserve University, Cleveland OH, 44106, USA}
\address{$^8$ Present Address: Jet Propulsion Lab, Pasadena, CA 91109, USA}
\address{$^9$ Present Address: Department of Physics and Astronomy, 
University of California, Los Angeles, CA 90095, USA}
\address{$^{10}$ Present Address: Department of Physics and Astronomy, 
University of Pennsylvania, Philadelphia, PA 19104, USA}
\address{$^{11}$ Present Address: Department of Physics,
Cornell University, Ithaca, NY 14853, USA}
\address{$^{12}$ Present Address: Laboratoire de Physique Corpusculaire
et Cosmologie, Coll\`{e}ge de France, F-75231 Paris CEDEX 05, France}

\newpage 

\begin{abstract}

We describe the design and performance 
of the Solar Tower Atmospheric Cherenkov Effect Experiment
detector in its initial configuration (STACEE-32).
STACEE is a new ground-based gamma ray detector
using the atmospheric Cherenkov technique.
In STACEE, the heliostats of a solar energy research array are
used to collect and focus the Cherenkov photons produced 
in gamma-ray induced air showers.
The large Cherenkov photon collection area of STACEE
results in a gamma-ray  energy threshold
below that of previous detectors.

\end{abstract}
\end{frontmatter}

\section{Introduction}

This paper describes the initial configuration of
STACEE, a new ground-based gamma-ray detector.
Before addressing the detector itself we provide a brief
summary of the current experimental situation in gamma-ray
astronomy and show the motivation for STACEE and similar detectors.

Gamma-ray astronomy has recently become a very exciting 
area of research.
During the lifetime of NASA's Compton Gamma Ray Observatory (CGRO) from
April, 1991 to May, 2000 and
following the construction  of ground based detectors during the 1990's, 
the field experienced rapid growth.
The amount and quality of data increased and theoretical 
understanding of the related astrophysics improved greatly.

\subsection{Space-borne Instruments}
Two instruments that were aboard the CGRO are of special interest to 
high energy astrophysics.
The Burst and Transient Source Experiment (BATSE) amassed a large
data set of enigmatic gamma-ray bursts (GRBs) and the Energetic Gamma Ray 
Experiment Telescope (EGRET) produced a catalog of over 200 
high energy point sources \cite{hartman99}.
Six of these sources have been identified with pulsars within our galaxy
and more than 70 have been found to be active galactic nuclei (AGNs)
at great distances.  
Optical or radio counterparts for the remaining sources have yet to be 
identified.

The EGRET telescope detected gamma rays 
by converting gamma ray to $e^+e^-$ pairs in a spark chamber
tracking device surrounded by an anti-coincidence shield which vetoed
charged particles.
This latter feature ensured an excellent signal-to-background ratio. 
EGRET could operate in this way because it was in orbit above the earth's
atmosphere.
Thus it was necessarily a small detector and could only look at sources 
below about 10 GeV.
This upper limit was given by statistics; the exact value 
was defined by intrinsic source strength, the steepness of its spectrum
and the observing time allocated.

\subsection{Ground-based Detectors}

Most ground-based gamma-ray detectors use the 
atmospheric Cherenkov technique.
Most resemble the Whipple telescope \cite{weekes89}, 
which was the first experiment to convincingly detect gamma-ray sources.  
Typical Cherenkov telescopes 
detect gamma rays by using large, steerable mirrors to 
collect and focus Cherenkov light produced by the relativistic
electrons in air showers resulting from interactions of high energy gamma
rays in the upper atmosphere. 
This Cherenkov light is distributed on the ground in a circular
pool with a diameter of 200-300 m.
The Cherenkov telescopes need only capture a small part of the 
total pool to detect a gamma ray so the telescopes 
have very large effective collection areas relative to satellite detectors,
albeit at higher energy thresholds. 

The energy threshold for Cherenkov telescopes is the result of a 
competition between collecting a small number of photons from a low energy
shower and rejecting a large number of photons from night sky background.
It dictated by a number of important parameters, 
including photon collection area ($A$), detector field of view ($\Omega$), 
integration time ($\tau$)
and efficiency for getting a photoelectron from a photon
hitting the primary collection mirror ($\epsilon$).
It can be summarized in the following approximate formula:

$E_{th} ~ \sim ~ \sqrt {\frac{\Omega \tau}{A \epsilon}}$

All of the parameters except $A$ are limited by current technology
($eg$ $\epsilon$ depends partly on quantum efficiency of photocathodes)
or by the physics of the air shower ($eg$ $\tau$ cannot be less than 
the time over which shower photons arrive at the detector.)
The only parameter which is practical to control is the collection 
area of the instrument. 
For present generation imaging detectors, this is less than 100 m$^2$
and thresholds are typically greater than 300 GeV.

The energy range between EGRET and the Cherenkov telescopes  remained 
unexplored until recently
since there were no detectors sensitive to the region from 
10 GeV to 300 GeV.
There are, however, new detectors being built or commissioned that
use the solar tower concept.
This concept is a variant of the air Cherenkov technique whereby the 
collecting
mirror is synthesized by an array of large, steerable mirrors (heliostats) 
at a central-tower solar energy installation.
The large effective size of the collecting mirror allows one to trigger
at lower photon densities and therefore lower primary gamma-ray energies.

It should be noted that future satellite detectors such as GLAST \cite{glast}
will explore some of this region but will be statistics-limited
above some energy which will depend on the source strength. 
Ground-based detectors will be able to complement low energy
satellite measurements with 
data taken over shorter time intervals.
This is important for detecting flaring activity in AGNs as well as pulsed
emission from pulsars. 
With longer integration times they will be sensitive to fainter sources.

\subsection{The STACEE Project}

STACEE (Solar Tower Atmospheric Cherenkov Effect Experiment)
is designed to lower the threshold of ground based gamma-ray 
astronomy to approximately
50 GeV, near the upper limit of satellite detectors.
Three other projects of a similar nature, CELESTE \cite{giebels98},
\cite{smith99},
SOLAR-2  \cite{zweerink99} and GRAAL \cite{diaz} have also recently 
been built or are under construction.

\begin{figure}[h]
\centerline{
\psfig{figure=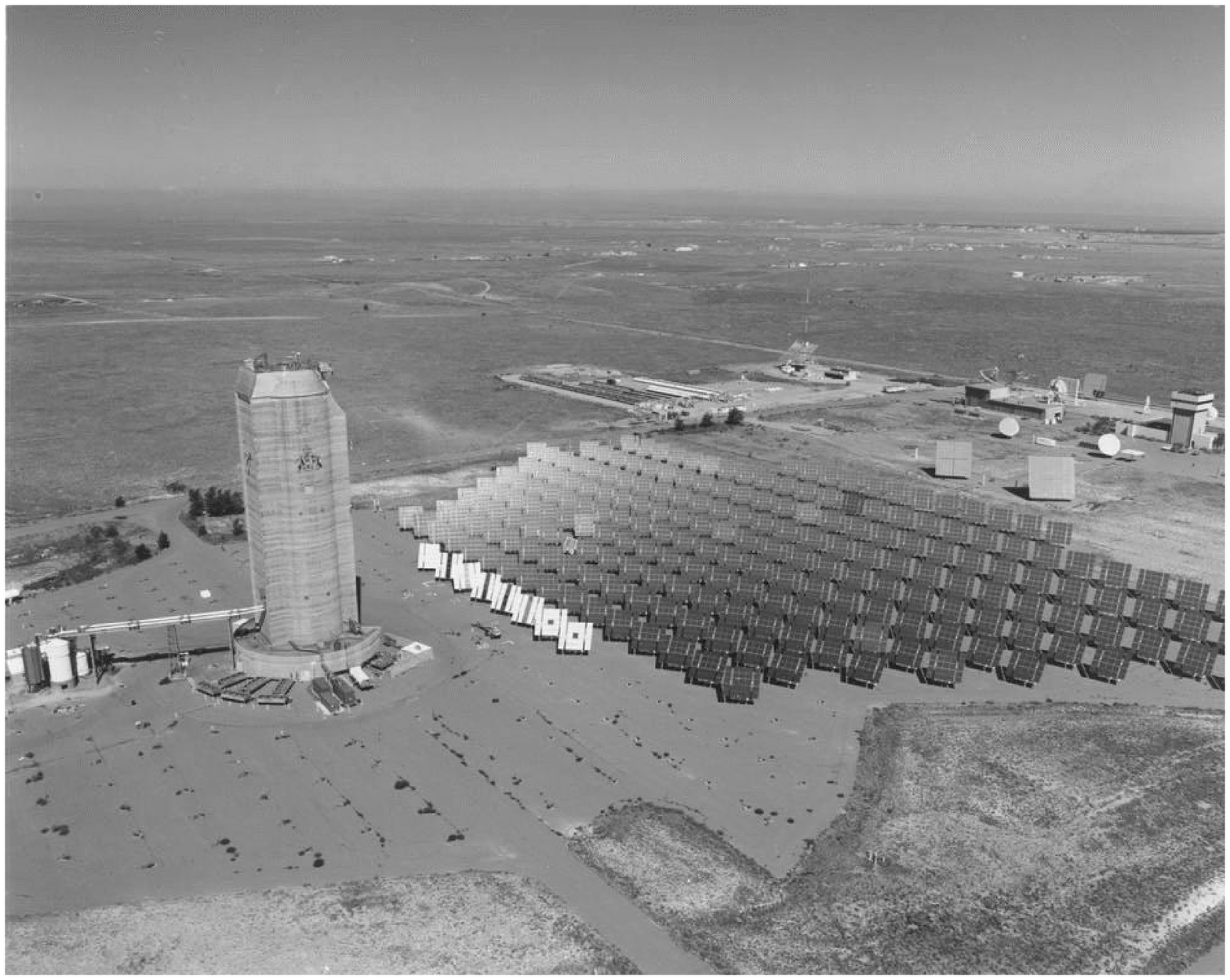,height=4.0in,bbllx=25pt,bblly=105pt,bburx=590pt,bbury=650pt,clip=.}}
\vspace*{1.5cm}
\caption{The National Solar Thermal Test Facility at Sandia National 
Laboratories
near Albuquerque, NM. 
The 212 heliostats each have an area of 37 m$^2$ and 
are laid out in rows running 
in an east-west direction. 
The 60 m high tower is located on the southern edge of the array.}

\label{nsttf}

\end{figure}

\begin{figure}[h]
\centerline{
\psfig{figure=
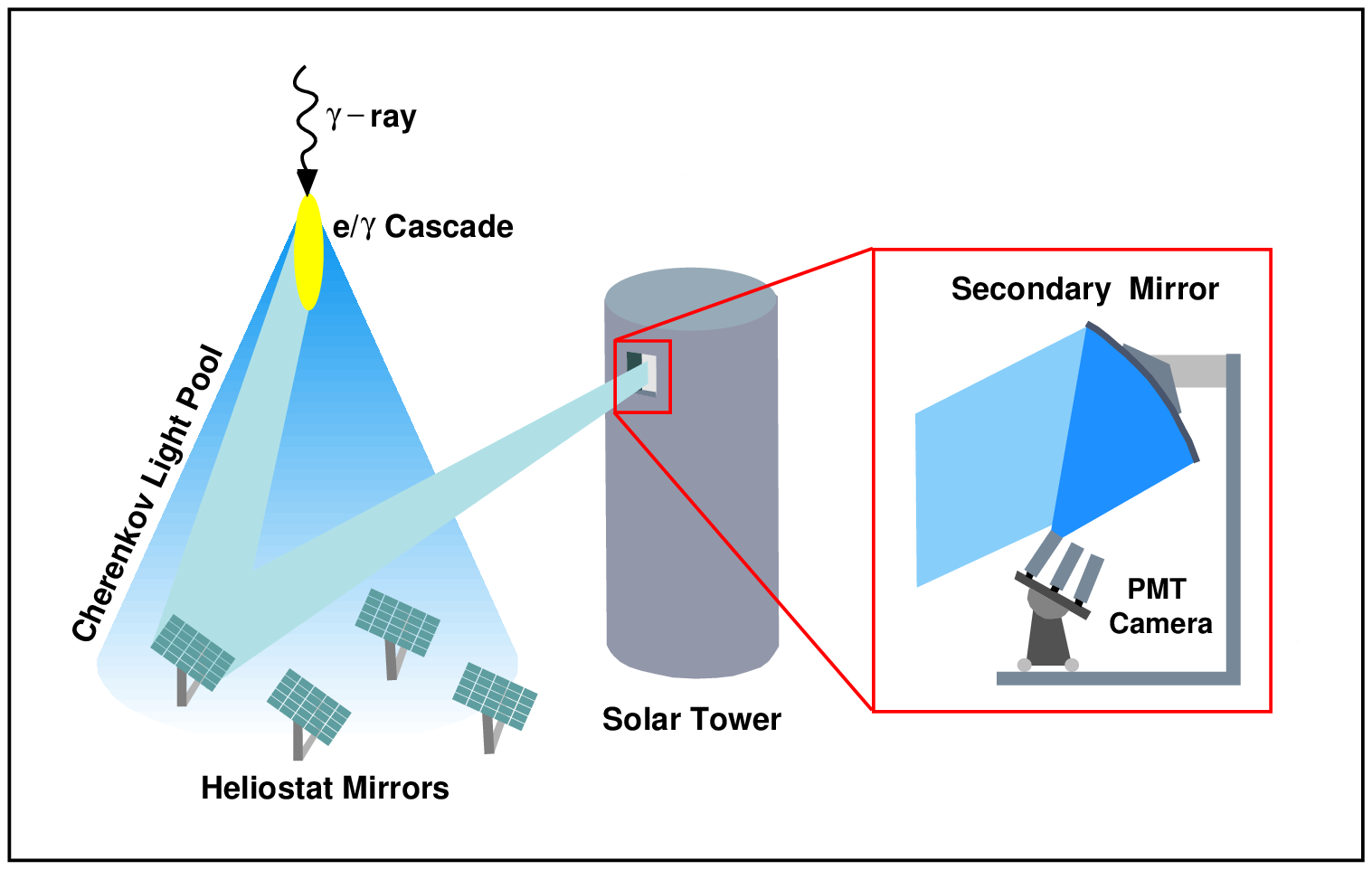,height=4.0in}}
\vspace*{0.5cm}
\caption{Concept of solar tower Cherenkov detection of gamma-ray 
air showers (not to scale):
Cherenkov light from a gamma-ray induced air shower (maximum
at about 10 km altitude) forms a pool of about 250 m diameter on the
ground. Some of the light is directed by large heliostats onto 
a smaller secondary mirror located at a height of 50 m  on the central tower.
The secondary focuses this light onto a matrix of photomultiplier tubes.
Each photomultiplier tube detects the light from a single heliostat.}

\label{stacee-concept}

\end{figure}

STACEE will investigate established and putative gamma-ray 
sources.
One of its principal aims is to follow the spectra of AGNs out to  
energies beyond that of EGRET measurements to determine 
where the spectra steepen drastically. 
These cut-offs are expected since, although Whipple-type
Cherenkov telescopes  
have the sensitivity to see many of the EGRET AGNs if their
spectra continue without a break, they have not detected them.
This effect could be due to cut-off mechanisms intrinsic to the source or to 
absorption effects between the source and the detector.
A likely mechanism is pair production wherein the high energy gamma ray
combines with a low energy photon (optical or infrared)
from the extragalactic background radiation field. 
The infrared (IR) component of this field is 
difficult to measure directly
but can provide information on the early universe since the photons are
red-shifted relics of light from early galaxies.
Correlating spectral cut-offs with source distance will help elucidate 
the nature of the IR field.

The field of high energy gamma-ray astronomy has recently been reviewed in
three comprehensive  articles \cite{ong98}, 
\cite{hoffmann99}, \cite{catanese}.

\section{STACEE-32}

The STACEE detector has been under development since 1994. 
STACEE-32 represents an intermediate step in the progress towards the 
design detector  
which is very similar in concept but has
twice as many heliostats and more sophisticated electronics for triggering
and pulse measurement.
Earlier steps in the development of STACEE are described in 
\cite{ong96} and \cite{chantell}.

STACEE-32 was installed at the National Thermal Solar Test Facility
(NSTTF), a solar energy research facility built by the US Department 
of Energy in the mid 1970's.
It is situated at Sandia National Laboratories in Albuquerque, New Mexico
(34.96$^o$~N, 106.51$^o$~W), at an altitude of 1700~m.
As shown in figure~\ref{nsttf}, the NSTTF consists of a 60 m concrete
tower and an array of 212 heliostats, each of which has a mirror 
area of 37 m$^2$.

Thirty two  
of these heliostats were used during clear, moonless nights to  
collect Cherenkov light from air
showers and direct it onto two secondary mirrors located near the top 
of the concrete tower.
The secondaries focussed the Cherenkov light onto banks of 
photomultiplier tubes
(PMTs) such that each PMT viewed a single heliostat.
Signals from the PMTs were amplified and routed to a control room where
they were processed by electronics and recorded if an appropriate trigger
condition was satisfied.
The essential concept is depicted in figure~\ref{stacee-concept}.

In the following subsections we describe in more detail the elements
of the detector.

\subsection{Heliostats}

\subsubsection{Location}

\begin{figure}[h]
\centerline{
\psfig{figure=
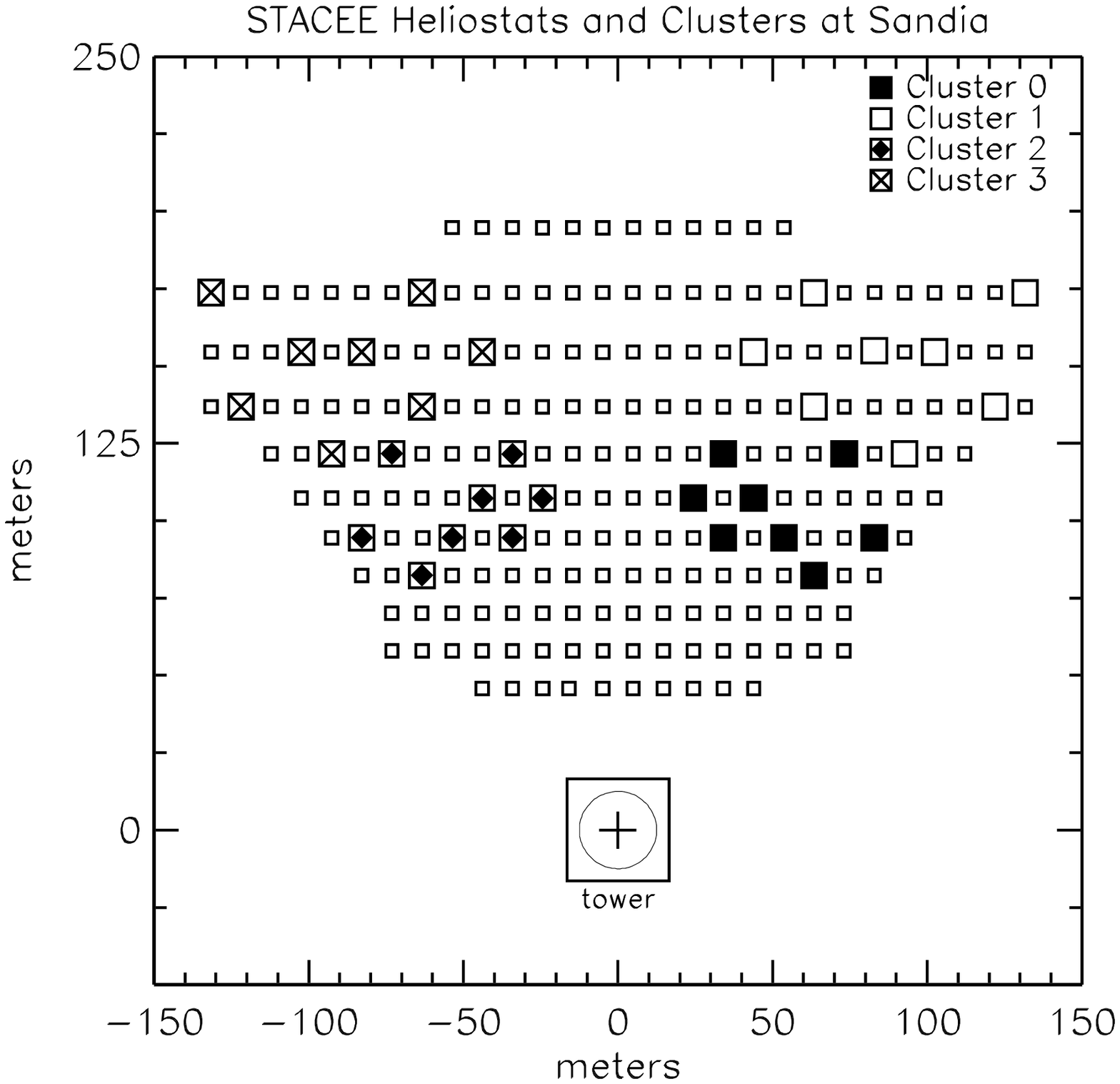,height=4.0in}}
\vspace*{1.0cm}
\caption{Plan view of the heliostat field showing the heliostats used for 
STACEE-32.
The 32 heliostats are grouped into 4 clusters for triggering purposes
as discussed in the text.}
\label{h-layout}
\end{figure}

The locations of the 32 heliostats used in STACEE-32, 
shown in figure~\ref{h-layout}, were chosen to sample
relatively uniformly the Cherenkov light pool
expected from a shower impacting near the center of the array. 
Crowding of PMTs in the image plane also played a role in the selection of 
heliostats.

\subsubsection{Mechanical Properties}

\begin{figure}[h]
\centerline{
\psfig{figure=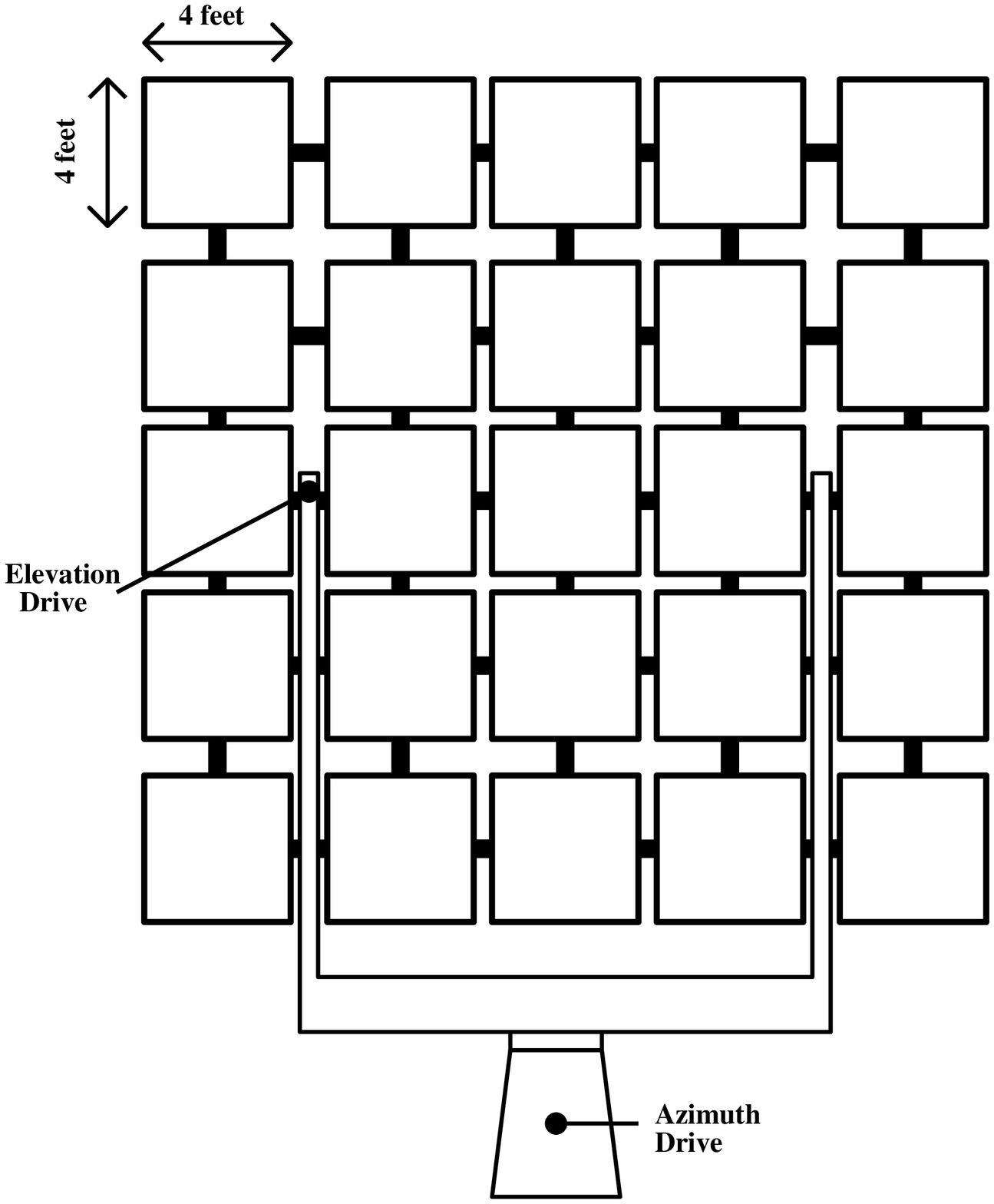,height=3.0in}}
\vspace*{0.5cm}
\caption{Schematic of a STACEE heliostat. Each of the 25 squares is a 
4 foot square of back-aluminized glass attached to a metal mounting
arrangement that is used to orient and focus it.}
\label{heliostat}
\end{figure}

Each heliostat consists of 25 square facets mounted on a steel framework,
as shown in figure~\ref{heliostat}. 
Each facet is a 4 foot by 4 foot 
square of back-surfaced aluminized glass glued
onto a thin metal sheet.
A metal suspension is welded to the back of the facet and bolts attached
to this suspension allow it to be fixed to the heliostat frame.
A bolt at the center of the facet and four at the corners are used to 
distort the facet into an approximately parabolic shape, the focus of 
which is set to be equal to the distance between the heliostat and the 
tower.
Other bolts precisely set the orientation of the facet with respect 
to the heliostat frame.
Each facet can be separately aligned so that their beams overlap 
at the tower.

The adjustment of the facet focusing 
was done during NSTTF commissioning. 
Since then the facet alignment was checked and tuned using images of the Sun
projected onto the tower near solar noon. 
A well tuned heliostat produced a small, bright ``sunspot'' composed
of 25 overlapping images of the Sun. 
Facets out of alignment caused distortions or broadening of this spot
and were brought into line using the image quality as a diagnostic.
Note that optimizing the image in this way does not assure that the 
optics are optimized for all locations of the Sun (or a Cherenkov
source) on the sky.
It is, however, a practical compromise.

We have tracked individual stars to check optical alignment. 
This provides some useful information but the sun is better as a 
diagnostic, partly because its angular size (0.5 degrees) is very 
similar to that of a Cherenkov shower. 
The point source nature of stars makes the procedure overly sensitive
to small errors in individual facet alignment or imperfections in 
their surfaces. 
An extended source like the sun smoothes these out and provides a 
better estimate of average alignment.

The entire heliostat is mounted on a Y-shaped yoke structure which
allows motion in azimuth and elevation.
This motion is achieved with two electric motors, each of which is
controlled by the NSTTF central computer using 13 bit encoders. 
The absolute pointing of the heliostats were determined using sunspots also.
The corresponding encoder readings, known as ``bias values'' were important
parameters in the operation of the heliostats.

\subsubsection{Optical Properties}

The important optical properties of the heliostats are their reflectivities 
and the size of their sun-spots.
Both of these directly affect how many Cherenkov photons will be 
directed onto a PMT.

\begin{figure}[h]
\centerline{
\psfig{figure=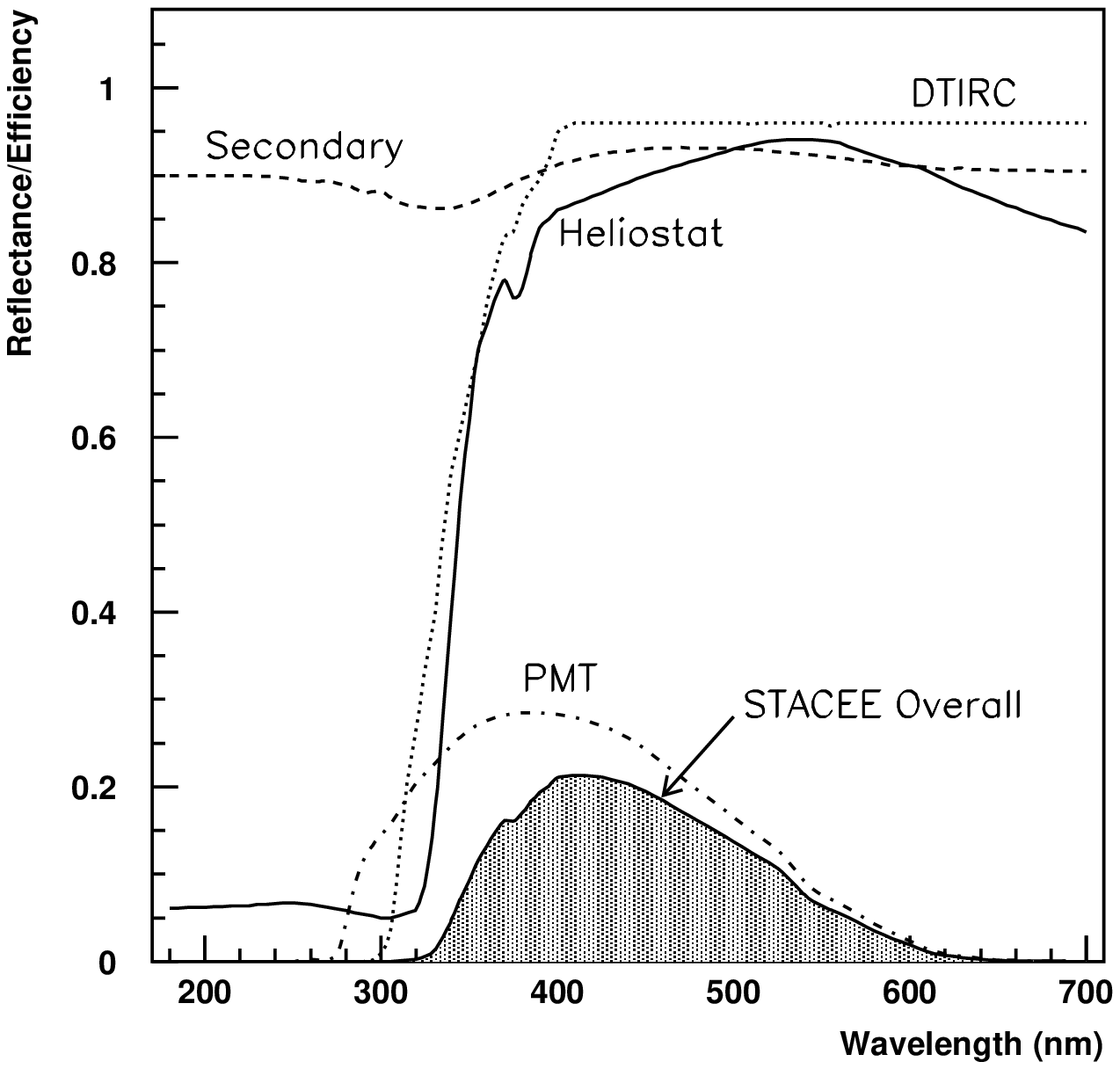,height=4.0in,bbllx=25pt,bblly=105pt,bburx=590pt,bbury=650pt,clip=.}}
\vspace*{-2.0cm}
\caption{Summary plot for the throughput of the STACEE-32 optical system.
Plotted as a function of wavelength are the reflectivities of the heliostats
(solid curve) and the secondary mirrors (dashed) 
as well as the transmittance of the DTIRC optical 
concentrators (dotted) and the quantum efficiency of the 
photocathode (dot-dashed).
The net effect of these components is displayed as the hatched region.}

\label{reflect}
\end{figure}

The heliostat facets are made from back-surfaced glass to 
withstand decades of weathering in a desert environment.
The double transit of light through the glass results in a cut-off in the
reflectance curve, as shown in figure~\ref{reflect}.

As previously mentioned, the focus and facet alignment of the heliostats
were checked and adjusted using the image of the Sun or Moon on the 
NSTTF tower. 
To quantify the size and brightness of each heliostat's sunspot, a
photograph was taken using a CCD camera and the data were analyzed 
to give information on the diameter of each spot and the amount of light
contained with circles of specified diameter.
This information was used in the simulations used to calculate the optical throughput of the 
detector.

The heliostats used for STACEE-32 were all inspected after being chosen 
and were found to be of uniformly high quality. 
Although no systematic tests were carried out on the individual facets,
visual inspection, sunspot images and performance in the experiment 
are consistent with the heliostats being approximately identical. 

\subsection{Secondary Mirrors}

Cherenkov photons were reflected by the heliostats onto two secondary mirrors 
located near the top of the central tower on a platform located 48 m
above the base of the tower.

One mirror viewed the 16 heliostats in the eastern part of the array while the 
other viewed the 16 heliostats in the west.

\subsubsection{Optics}

The secondary mirrors were spherical, with a nominal diameter of 
1.9 m and a focal length of 2.0 m. 
They focussed the light from the heliostats, which arrived as a wide beam, onto phototube
cannisters fixed in position at the focal plane.
The optics were such that each heliostat was mapped onto its corresponding PMT
channel.
This one-to-one mapping was vital for pattern recognition which was used in 
trigger formation and background suppression. 

The off-axis geometry and spherical optics gave rise to coma aberrations 
in the heliostat images. 
The coma caused a certain amount of cross-talk since light from one heliostat's tail
could end up in the wrong PMT. 
However, it was usually possible to reject this light because in almost all 
cases it arrived 
at the wrong time
as a result of the different time-of-flight from its heliostat.
For certain pointing geometries, 
cross-talk could arrive close enough in time to be
included in the ADC gate but this would not significantly degrade the 
analog measurement which was already quite contaminated with night sky 
background integrated during the long (37 ns) gate.
To degrade the timing measurement, cross-talk photons would need to arrive
closer in time to be accepted by the trigger and by offline cuts.
On any given event only one or two channels would be affected.
Typically, the coma tails contained less than 10\% of a channel's 
light so the net effect of timing and flux levels meant that cross-talk
was not an important effect.

Each secondary mirror was composed of seven identical hexagonal facets made from
front-surfaced aluminized glass covered with a silicon oxide protective coating.
The reflectance curve, as provided by the manufacturer \cite{gmo},
is shown in figure~\ref{reflect}.

\subsubsection{Mechanics}

\begin{figure}[h]
\centerline{
\psfig{figure=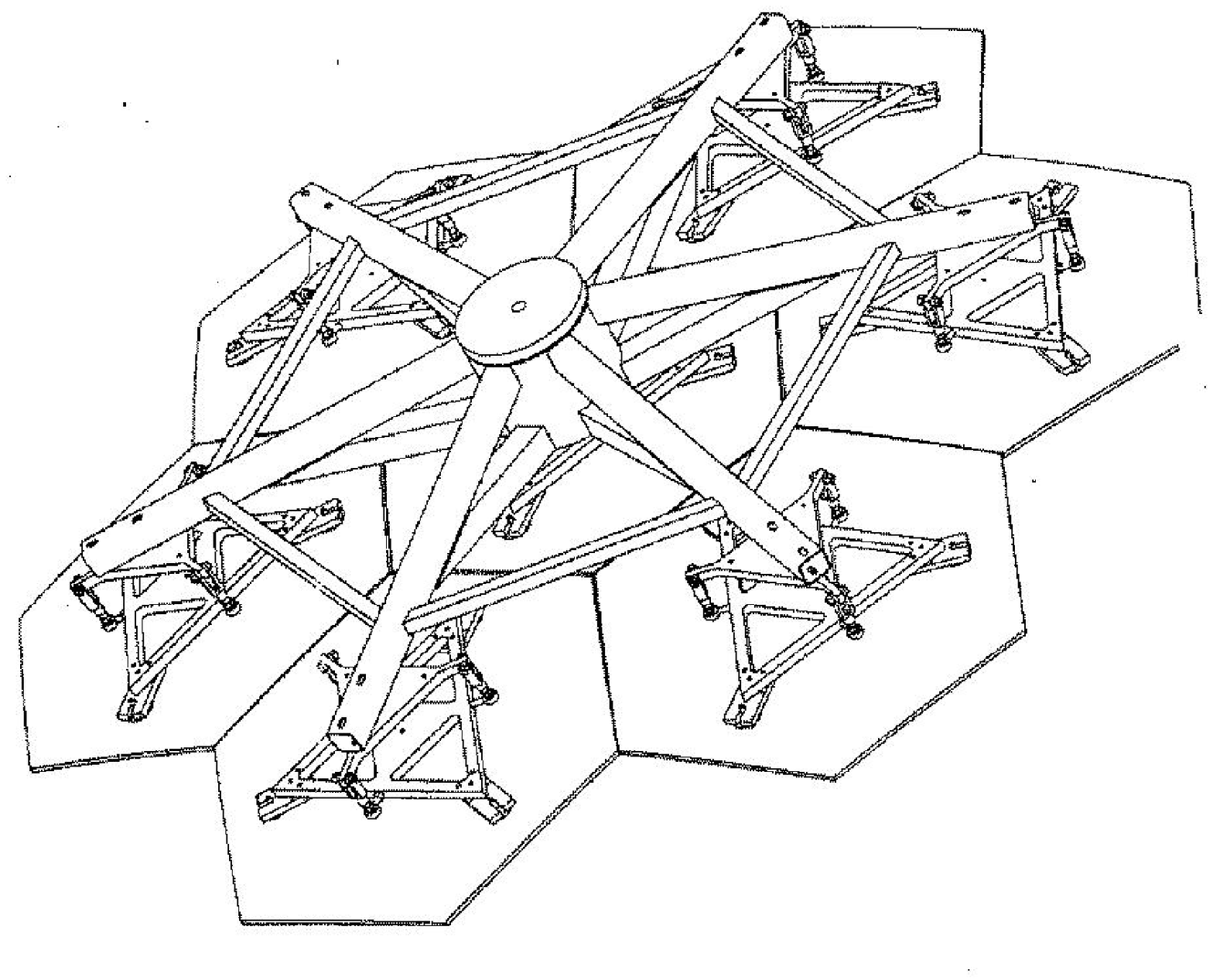,height=3.0in}}
\vspace*{0.5cm}
\caption{Drawing of a secondary mirror, detailing the suspension of the 
seven individual facets.
Each facet has three threaded studs glued to its back surface. 
The studs are attached to an aluminum triangle which  is itself attached
to a spider.
Except for the central facet, the attachment of each facet to the spider
was done with a triplet of turnbuckles which allowed the 
orientation of the facet to be precisely adjusted.
The approximate diameter of the entire assembly is 2 m.}

\label{spider}
\end{figure}

The facets were mounted on a spider, as shown in figure~\ref{spider}, with the 
center facet fixed in position and the outer six adjustable. 
The outer six were aligned by placing a point source of light,
surrounded by a small white screen, at the center of curvature of the mirror and 
adjusting the facets until all reflected images overlapped.
With the facets aligned the resultant spot was approximately
1 mm in size.

The spider itself was attached at the pivot point of a jib crane mounted on a stationary 
forklift, or stacker.
The stacker allowed the raising of the mirror from its daytime stow position to its
nighttime operational position. 
Jack screw actuators connecting the spider and the jib crane assembly  controlled the 
pitch and yaw orientation of the mirror and were set once, upon installation.
The axes of the secondaries made an angle of $32^o$  with respect to the horizontal.

\subsection{Camera}

The final stage in the STACEE optics chain is the camera. 
In STACEE-32 there were two cameras, one for each secondary mirror.

\begin{figure}[h]
\centerline{
\psfig{figure=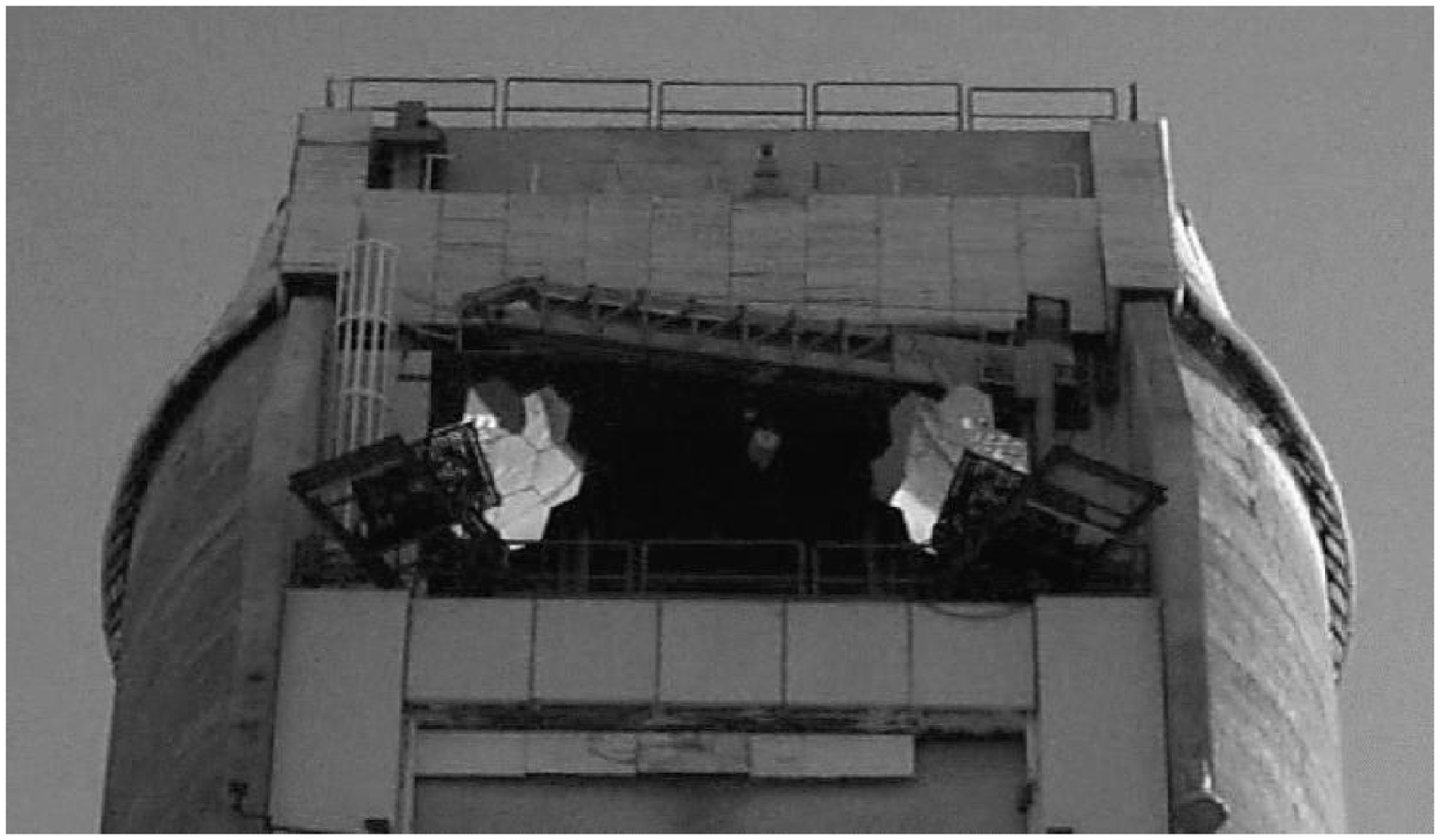,height=3.0in}}
\vspace*{1.0cm}
\caption{A photograph of the secondary mirrors and cameras 
used in STACEE-32.
The mirrors are visible as the structures with seven hexagonal facets.
The structures which protrude from the tower and partially occult the
secondaries are the PMT cameras and the skids on which they travel.}

\label{2mirrors}
\end{figure}

\begin{figure}[h]
\centerline{
\psfig{figure=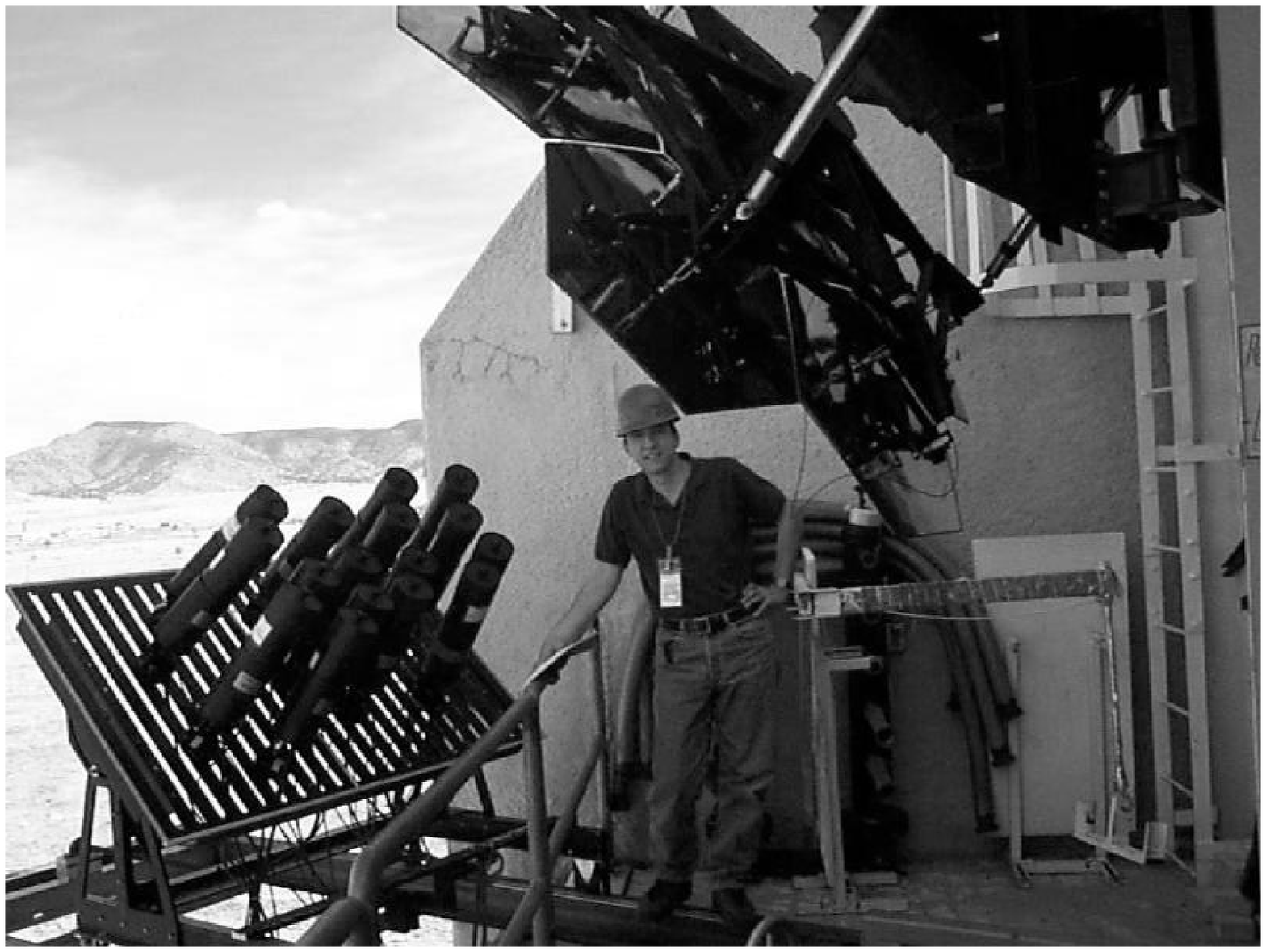,height=4.0in}}
\vspace*{0.5cm}
\caption{Photograph of the east camera and secondary
of STACEE-32,
along with one of the STACEE collaborators.}

\label{rene_photo}
\end{figure}

A photograph of the mirrors is shown in figure~\ref{2mirrors}
and a photograph of one of the cameras and mirror assemblies  is shown in 
figure~\ref{rene_photo}.

\subsubsection{Mechanical Details}

Each camera consisted of 16 PMTs and light concentrators enclosed in 
cannisters mounted on a slotted aluminum plate attached to a trolley.
The trolley rolled on clamping wheels, controlled by a threaded rod  
along a skid made from welded steel I-beams.
The skid extended from under the stacker and ran parallel with the 
horizontal component of the secondary mirror's axis. 
About two thirds of the skid rested on the platform of the solar tower
and the remaining third extended out from the tower.
The trolley was rolled in for stowage and out for observing. 

Mounted on the trolley was a 1.0 m x  1.7 m aluminum plate machined 
with 25 mm slots on a 75 mm pitch.
This plate swivelled in elevation and was set to $44.8^o$ with respect to 
vertical. 
This angle was prescribed by the optics of the system in the following
way. 
A line from the center of the group of heliostats observed by the camera
to the center of the secondary mirror defines the primary optical path.
This line should reflect from the secondary to the center of the camera
when it is in its correct position on the skid.
That position is defined
by the focal length of the secondary mirror. 
The tilt angle of the slotted plate is defined by requiring its normal
to point to the center of the secondary. 
The plate was tilted to this angle using a digital protractor and secured
using jam nuts.

\begin{figure}[h]
\centerline{
\psfig{figure=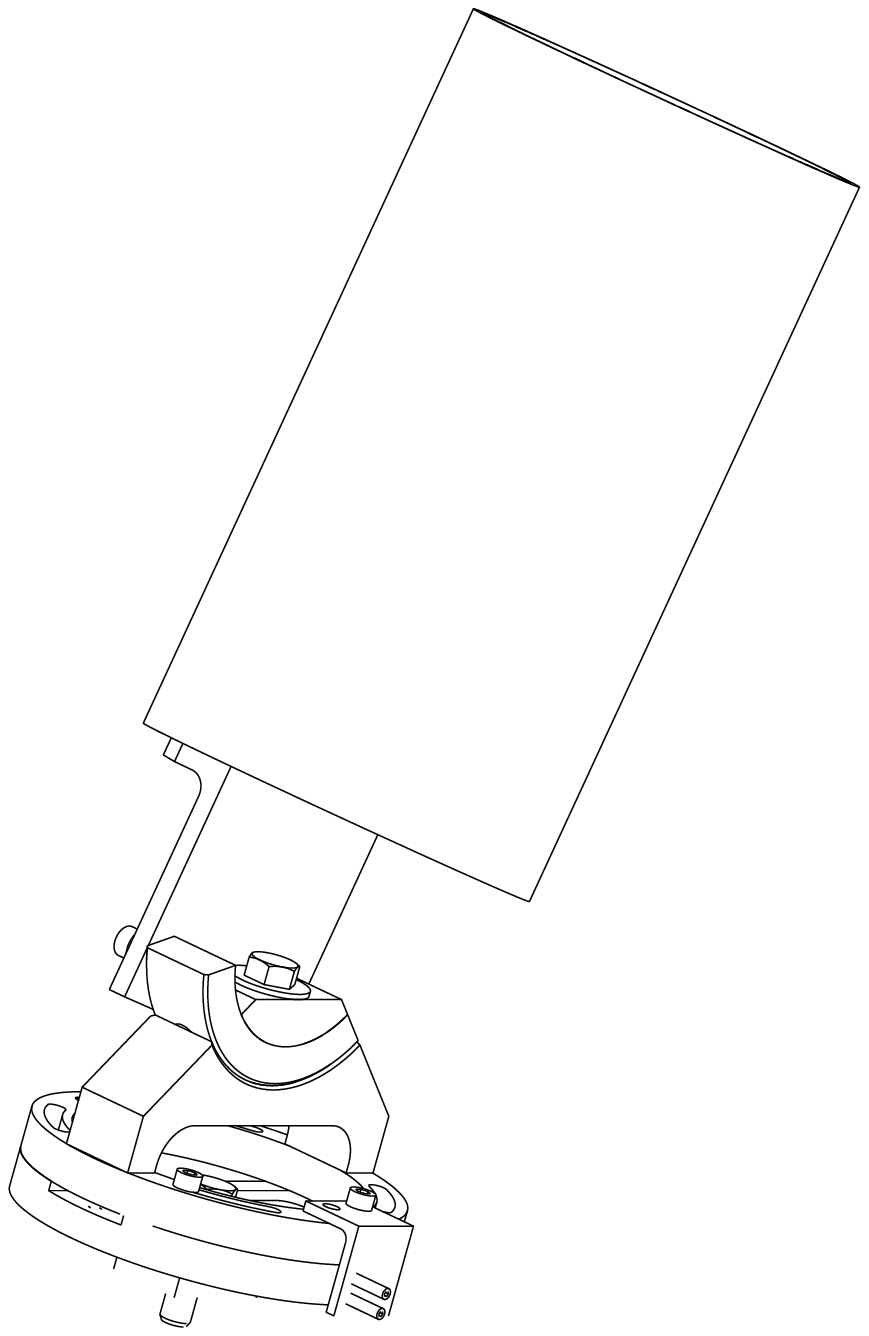,height=3.0in}}
\vspace*{0.5cm}
\caption{Drawing of the azimuth-elevation mounting system which allows 
each phototube
cannister to be precisely positioned and oriented on the camera.
Shown is the sleeve, into which the PMT cannister slides, connected 
by a bracket to a metal arc which allows tilting in elevation.
This assembly is mounted on a turntable for azimuthal adjustment.} 

\label{az-el}
\end{figure}

The PMT cannisters were mounted in cylindrical sleeves attached to 
an azimuth-elevation  
mounting system secured to the slotted plate.
A drawing of this assembly is shown in figure~\ref{az-el}.
With this system it was possible to position a PMT cannister anywhere
laterally on the slotted plate and to adjust the orientation of the cannisters
such that they pointed to the center of the secondary mirror.
Shims in the bottom of the sleeves provided adjustment along the axis of 
the cannister. 
This longitudinal degree of freedom was necessary to accommodate the 
curved focal plane of the system.

\begin{figure}[h]
\centerline{
\psfig{figure=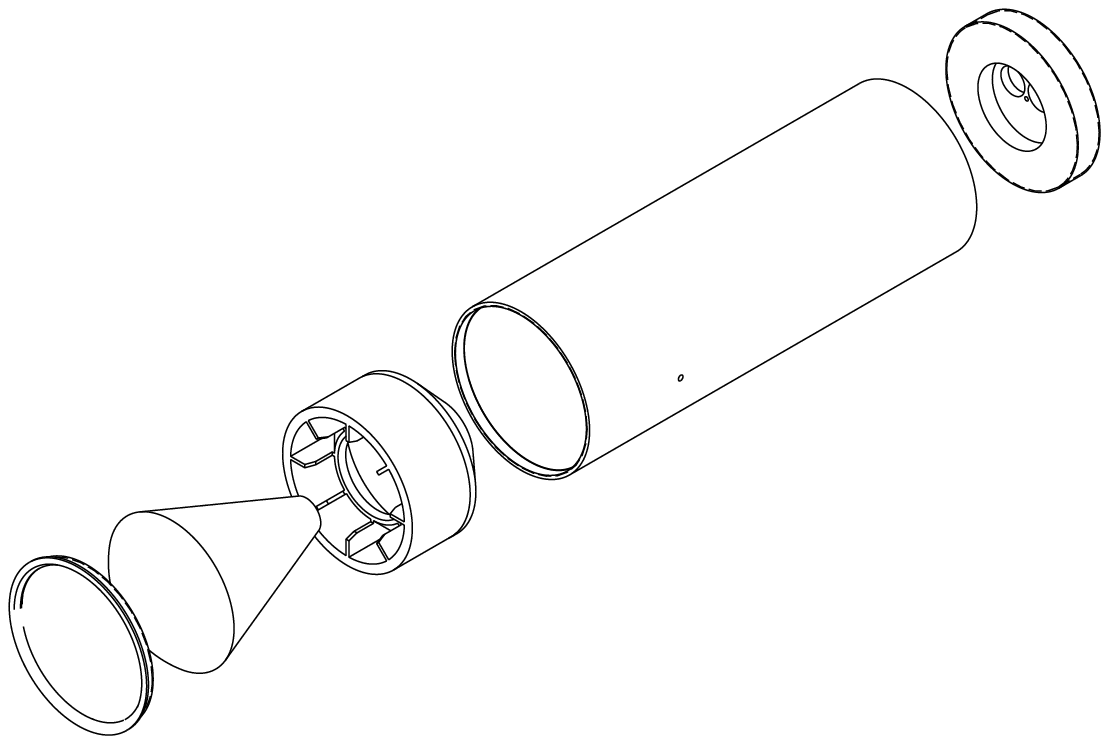,height=3.0in}}
\vspace*{0.5cm}
\caption{Exploded view of a PMT cannister and its contents. 
From the left, the components shown are: a threaded retaining ring,
a DTIRC (light concentrator), a locator device to hold the DTIRC
in position, the cannister and its end-disk.   
Not shown is the PMT, which attaches to the small recess in the end-disk.}

\label{cannister}
\end{figure}

An exploded view of a PMT cannister is shown in figure~\ref{cannister}.
It contained the PMT (not shown) a silicone rubber optical coupling 
disk (also not shown) and a light concentrator, known as a DTIRC.
The PMT base was fitted into the recess in the disk at the back of the 
cannister tube with its HV and signal connectors protruding through holes
provided. 
The 3 mm thick silicone disk was sandwiched between the PMT and the 
DTIRC which was held in place by a locator device.
The entire assembly
was held in longitudinal compression by a thin ring screwed in from 
the front.

\subsubsection{Light Concentrators}

\begin{figure}[p]
\psfig{file=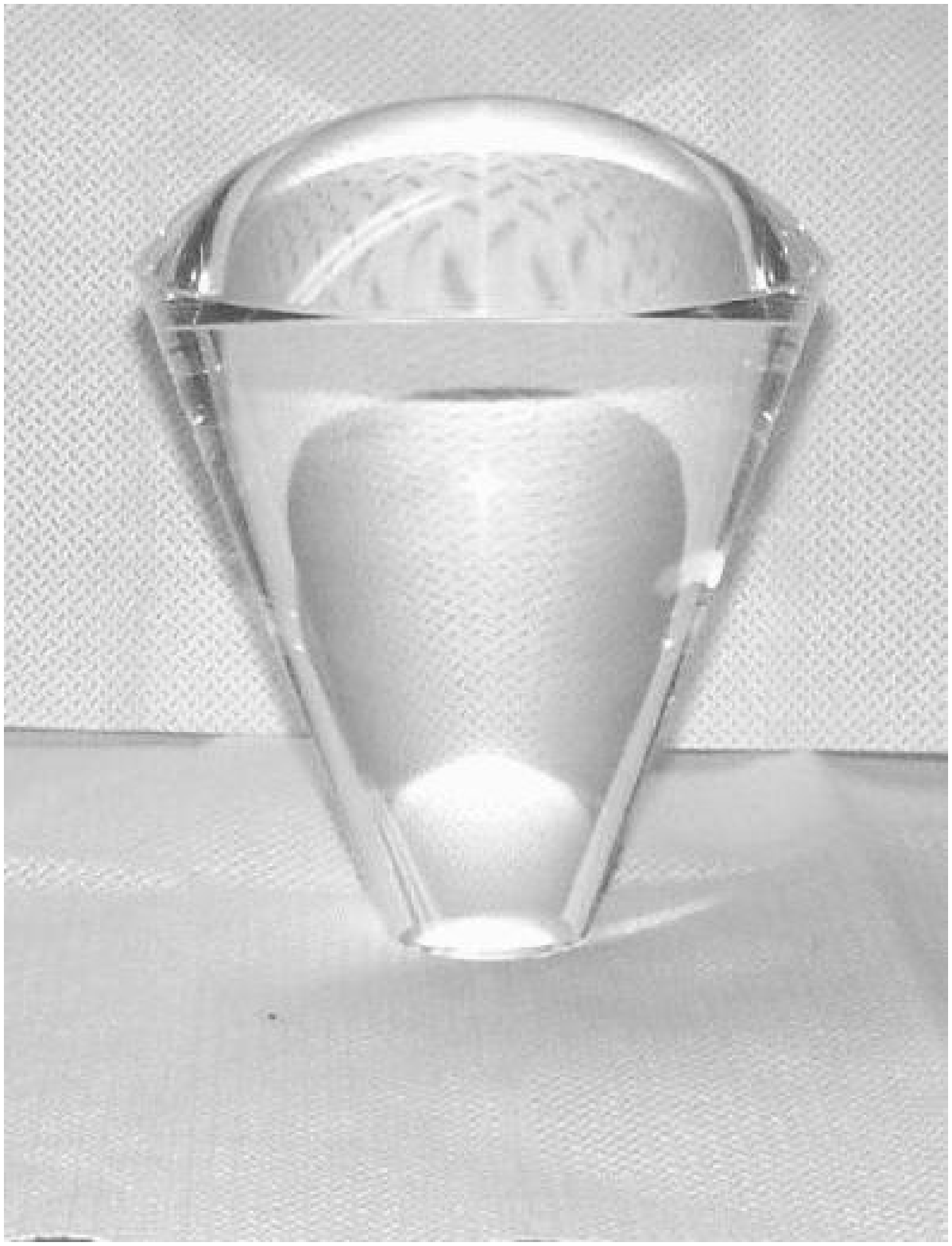,width=2.9in}
\epsfig{file=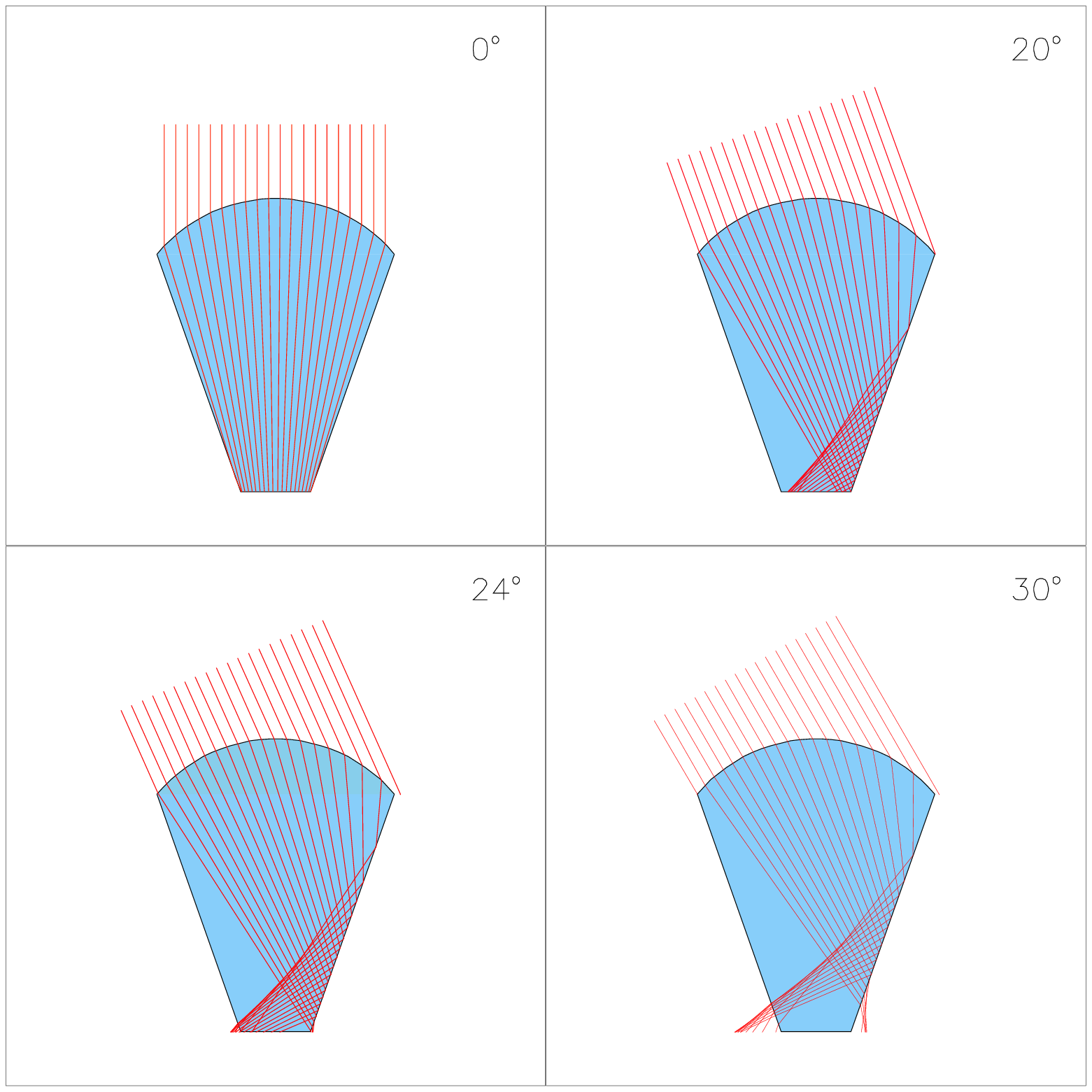,width=3.9in}
\caption{
Dielectric Total Internal Reflecting Concentrator (DTIRC).  
{\em Upper:} A photograph of a STACEE DTIRC.  
{\em Lower:} A DTIRC only accepts light
from within a fixed field of view. Within the acceptance angle, rays
reflect internally and proceed to the exit aperture. Outside the acceptance
angle, they are lost when hitting the sides of the DTIRC.
}
\label{fig.dtirc}
\end{figure}

The light concentrators used in STACEE were Dielectric Total Internal Reflection 
Concentrators (DTIRCs)
\cite{ning87}.
These are non-imaging devices which use total internal reflection to transport light 
from the front surface to the exit aperture;
the light from a circular area of 11 cm diameter was focussed to 
an exit diameter of less than 4 cm. 
Only light from a given angular range could reach the exit aperture
so the DTIRCs had the added 
feature of being able to define the field of view of the PMT.
Functionally, a DTIRC is very much like the more familiar Winston Cone
\cite{hinterberger66},
used extensively in particle physics detectors. 
In fact, it is a generalization. 
The DTIRC characteristics, defined by the shape of the front surface and 
the side profile,
can be varied over a range of possibilities. 
For compactness and ease of manufacture the STACEE design incorporated a spherical
front face and straight, tapering sides.
A photograph of a STACEE DTIRC is shown in figure~\ref{fig.dtirc}.

%\begin{center}
%\end{center}

\begin{figure}[h]
\centerline{
\psfig{figure=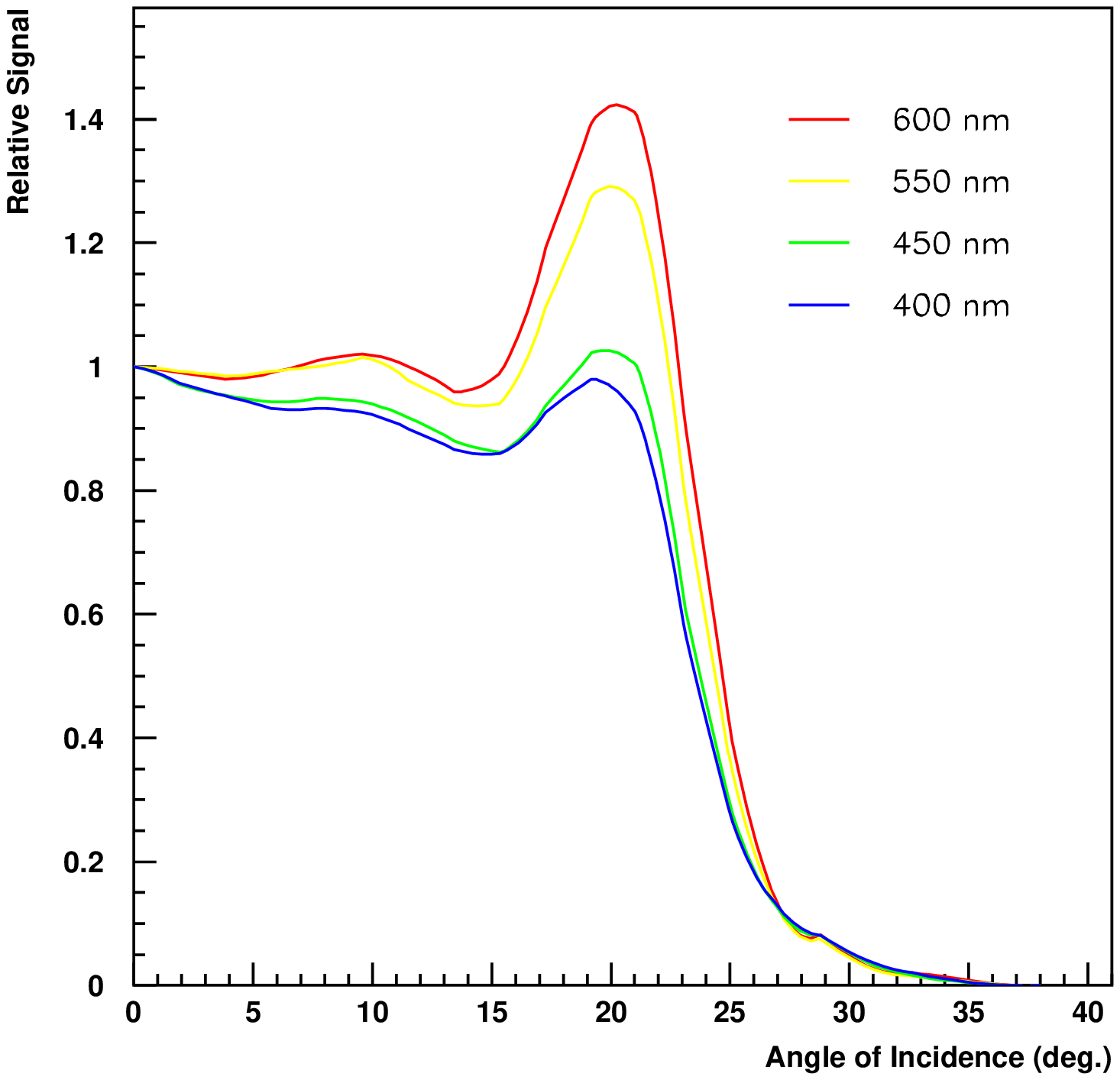,height=4.0in,bbllx=25pt,bblly=105pt,bburx=590pt,bbury=650pt,clip=.}}
\vspace*{-2.0cm}
\caption{Typical angular response of a
DTIRC/PMT assembly.
The PMT current is plotted as a function of incident angle of a 
parallel beam of light which illuminates the entire entrance 
aperture of the DTIRC.
The wavelength of the light for each of the four scans 
was selected using filters.
All scans have been scaled to unity at normal incidence. 
Note the cut-off in acceptance for angles above $23^o$; 
the DTIRC provides an effective way of controlling the 
field of view of a PMT.
The enhancement seen for angles near the cut-off
is largely due to photons impacting the photocathode at small 
angles.
The enhancement is more important at longer wavelengths.}

\label{enhance}
\end{figure}

The DTIRCs were made from solid UV-transparent acrylic manufactured for the Sudbury 
Neutrino Observatory \cite{boger00}. 
This material has transmission well into the ultraviolet (attenuation length 
greater than 10 cm for wavelengths greater than 320 nm) and has a refractive index of 1.49.
The transmittance $vs$ wavelength for a typical pathlength in a DTIRC is
shown in figure~\ref{reflect}.

Three types of DTIRCs were used, each with the same front aperture but differing cone 
angles and therefore different acceptance angles.
These were used to roughly equalize the fields of view of different channels.
The field of view of a given heliostat is given by the effective diameter of the secondary
mirror divided by the distance from the secondary to the heliostat. 
To equalize the fields of view we used DTIRCS with a narrow angular acceptance 
(which therefore view a smaller part of the secondary) to view nearby heliostats.
Large acceptance DTIRCs were used to view more distant heliostats.
For practical reasons, we limited the number of different DTIRC sizes to three rather than
make a different size for each channel.
Table~\ref{tab.dtircs} summarizes the properties of the three DTIRC types.

\begin{table}
\begin{center}  
\begin{tabular}{|llllll|}  
\hline
Full Angle  & Entrance & Exit     & Total  & Diameter     & Mean Distance\\
Acceptance  & Diameter & Diameter & Height & Viewed On    & to Heliostats\\
($^o$)      & (cm)     & (cm)     & (cm)   & Secondary (m)& (m) \\
\hline
$19.0$ & $11.0$ & $2.55$ & $14.40$ & $1.30$ & $116$\\
$24.0$ & $11.0$ & $3.25$ & $13.58$ & $1.63$ & $150$\\
$28.1$ & $11.0$ & $3.83$ & $12.96$ & $1.88$ & $194$\\
\hline
\end{tabular}
\end{center}
\medskip
\caption[DTIRC properties]{DTIRC properties. \label{tab.dtircs}
}
\end{table}

The DTIRCs all had the same entrance diameter. 
In the absence of optical aberrations it would have made sense to use smaller diameters 
for more distant heliostats since the size of their images on the focal plane was smaller.
However, these heliostats were further off axis than were the
nearby ones and therefore suffered 
more from the effects of coma. 
Thus, large diameter DTIRCs were necessary to capture 
more of this light.

After manufacture, the DTIRC angular responses were checked by performing
an angular scan. 
A wide, parallel beam of light was directed onto the front face of 
a PMT-DTIRC assembly and the resulting photocurrent was recorded as a function
of the angle of incidence.
A typical result is shown in figure~\ref{enhance}. 
The cut-off angle is as expected.
The rise in photocurrent for angles near the cut-off is the result
of photons impacting the photocathode at small angles. 
At such angles the photocathode is effectively thicker so the quantum
efficiency is enhanced.

\subsection{Photomultiplier Tubes}

STACEE required photomultiplier tubes (PMTs) with good sensitivity to 
short wavelengths (blue and UV) where most of the Cherenkov light is concentrated.
Each PMT viewed the light from a 37~m$^2$ heliostat
so it generated single photoelectrons from night sky background at a
rate in excess of 1.5 GHz.
(The precise rate was channel dependent, depending on the position 
of the heliostat which affected the field of view and the albedo.
This was taken into account in detector simulations.)
To reduce pulse pile-up effects, a PMT with a rapid rise time and narrow
output pulse width was required.
A small transit time spread was also desired since it resulted in better time resolution. 
The excellent time resolution allowed us to 
exploit the narrowness of the Cherenkov 
wave-front at the 
trigger level to reject background from showers produced by charged cosmic
rays.
Offline, good time resolution was valuable in reconstructing the shape of the wave
front (approximately spherical) in order to reject background.

\subsubsection{Choice of PMT}

The PMT used was the Photonis XP2282B with a borosilicate window and the  
VD182K/C transistorized voltage divider.
This tube's Sb-K-Cs bialkali photocathode provided a  peak quantum efficiency 
of $\sim 28\%$ at $\lambda = 400~$nm, and the borosilicate glass transmitted
UV light down to $\lambda = 280~$nm.
The XP2282B is a 51 mm  diameter tube with 8 linearly focussed dynode stages.
Under typical operating conditions it had a rise time of 1.5 ns and a transit
time spread of 0.5 ns.

Additionally, the XP2282B is rated for photocurrents of up to $180~\mu$A
which was important in our application.
The transistorized voltage divider provided good linearity by keeping the high voltage
distributed to the dynodes constant, independent of current drawn. 

Important tube parameters are summarized in table~\ref{2282B}.

The DTIRCs and phototubes were studied ~\cite{fortin}, with
LED, laser and broad-band light beams to check for efficiency as a function 
of position and angle of the incident photons.
The results were parametrized and included in the detector simulations.

\begin{table}
\begin{center}  
\begin{tabular}{|ll|}  
\hline
Tube diameter & 51 mm \\
Number of dynodes & 8\\
Typical Operating Gain	& $1.1 \times 10^5$ \\
Transit Time Spread & 0.5 ns \\
Typical Operating Voltage & -1600 V \\
Rise time & 1.5 ns \\
\hline
\end{tabular}
\end{center}
\caption[Properties of the XP2282B phototube]
{Properties of the XP2282B phototube.  Also shown are typical
operating parameters for STACEE-32. \label{2282B}
}
\end{table}

\subsubsection{High Voltage}

The PMTs were supplied with high voltage from a LeCroy 4032A high voltage 
supply located on the detector level and controlled by a LeCroy 2132
CAMAC interface in the control room via a long ribbon cable. 
Voltages were typically in the neighborhood of -1600 V.

\subsubsection{Gain Calibration}

\begin{figure}[h]
\centerline{
\psfig{figure=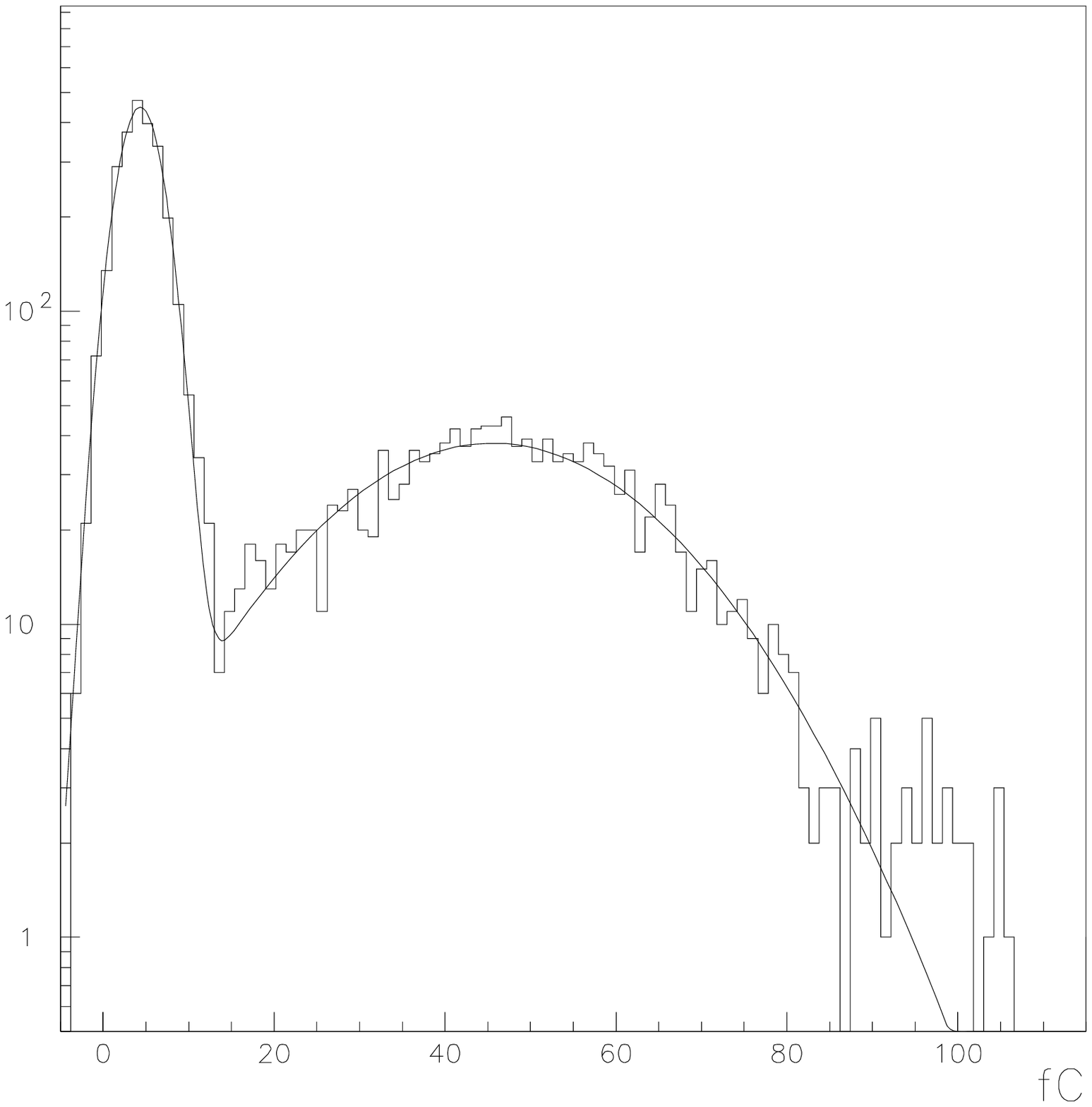,height=4.0in}}
\vspace*{0.5cm}
\caption{A typical pulse charge spectrum from a STACEE PMT held
under high voltage in a dark box. The mean number of photons striking 
the photocathode is much less than one. The data are fit to two 
Gaussian distributions, one for the pedestal and the other for the single
photoelectron peak.
Spectra like these were used in determining the absolute gain of each 
PMT.}

\label{singlepe}
\end{figure}

The operating voltages for the PMTs were determined by measuring the 
gain of each channel at several high voltage settings. 
These measurements were fit to the function:
\begin{center} 
$ G \propto V^\gamma$ 
\end{center} 
and the function was inverted 
to get the voltage required to obtain the desired gain. 
To measure the absolute gain for a PMT, we placed it in a dark box and 
obtained a spectrum of single photoeletron pulses using a 2 GS/s digital 
oscilloscope in self-triggered mode, adding up the samples over a 
20 ns range. 
A sample pulse height spectrum is shown in figure~\ref{singlepe}.

The gains were adjusted slightly, 
after installation of the PMTs in the detector,
using a collimated LED to produce light pulses 
(approximately 50 photo-electrons each) in each
PMT separately.
The high voltage values were changed to equalize the 
response of all channels to the LED signal.

\subsection{Electronics}

\subsubsection{Overview}

\begin{figure}[h]
\centerline{
\psfig{figure=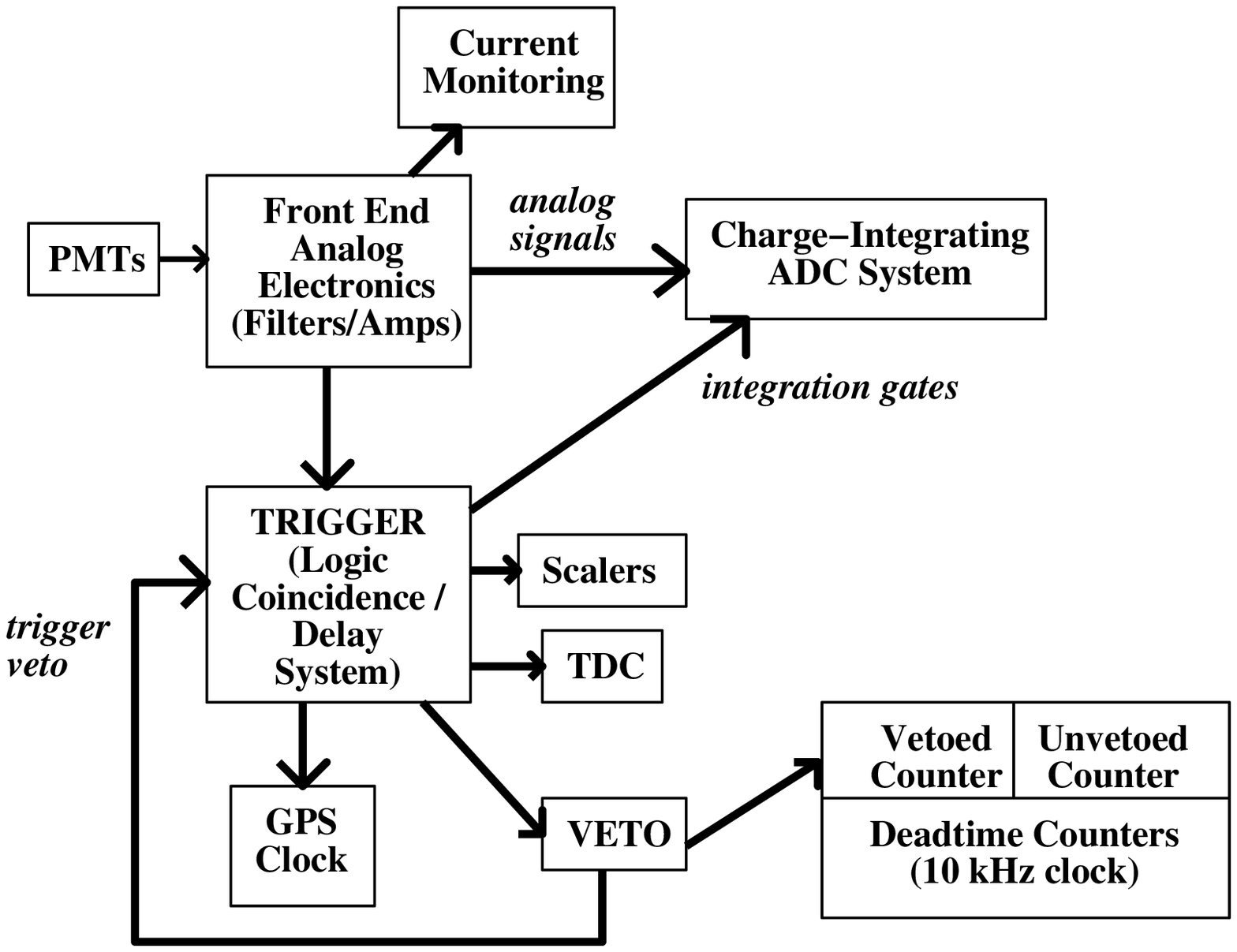,height=3.0in}}
\vspace*{0.5cm}
\caption{Block diagram of the STACEE-32 electronics. 
The front end analog electronics filters and amplifies (x100) the phototube signals.
The trigger system combines the signals in a high-multiplicity coincidence to form a trigger.
The trigger signal latches the GPS clock time, 
asserts a veto, stops the TDC counters and sends 
integration gates to the ADC system.
Deadtime counters record the deadtime fraction of the experiment.}

\label{electronics}
\end{figure}

The STACEE-32 electronics could be grouped into subsystems corresponding 
to their function in the experiment, as shown
in figure~\ref{electronics}.
Signals from the phototubes were filtered and amplified near the cameras 
then sent to the STACEE control room, located 20 m below the detector in the
tower.
Here they were discriminated and used in trigger 
logic and timing measurements. 
Analog pulses from selected channels were digitized.
  
The electronics were an assembly of commercially available NIM and CAMAC 
modules
which served to define a trigger and measure the times and charges of
pulses from the PMTs. 
As such, the electronics were very similar to those found in a typical high-energy 
physics experiment.  

A major difference, however, was the requirement of a dynamic delay system.
Due to the earth's rotation, the gamma-ray source appeared to move across
the sky during the course of a night's observations. 
This effect continuously changed 
times of arrival of Cherenkov photons at each heliostat.
In order to maintain tight coincidences, signals from different channels
were required to  be delayed by amounts which had to be 
incremented in nanosecond steps every few seconds.
The range of these delays was of order 250 ns. 
This was one of the design challenges faced and it influenced the configuration 
presented here.

\subsubsection{Front-End Analog Electronics}

The front-end analog electronics were physically close to the 
PMTs;
they were installed in a small hut next to the cameras on the detector 
level of the NSTTF tower.
The PMT signals arrived at the hut via 11 m long RG58 cables which were 
bundled, in groups of 4, inside ground braids of the same length.
In the hut, the signals passed through a 
high-pass RC filter having a time constant of 75 ns.
This filter blocked any DC component of the PMT signal and removed
slow PMT transients which were not associated with Cherenkov signals. 
The DC component of the photocurrent developed a voltage across the 
resistor in the filter 
and this voltage 
was sent by ribbon cable to a Joerger ADC-32 scanning ADC module in the 
control room.
Typical currents were between 20 and 40 $\mu$A.

The pulsed components of the signals exiting the filters were amplified by 
two cascaded fixed gain (x10) wide-band (275 MHz) amplifiers (Phillips Scientific 776).
This amplification 
factor of 100 allowed us to keep the PMT gain to $\sim~10^5$ which 
prolonged the life of the PMTs in an environment of high night sky background light levels.

The filtered and amplified signals were routed through 40 m low-loss coaxial
cables (RG213) from the detector level to the control room level of the tower
where they were fed into
linear fanouts (Phillips Scientific 748).
The outputs of these fanouts were passed to the discriminators and ADCs.

The effect of these components on the PMT pulse was to broaden and
attenuate it.
The electronics had a negligable effect compared with the cables.
A typical pulse was about 4 ns wide at the PMT and 6 ns wide after the 
40 m cables. 
(Additional cables were used to delay the pulses which were sent to 
the ADCs. 
These are described later; their effect was to broaden the pulses to 
approximately 25 ns.)

\subsubsection{Trigger Electronics}

\begin{figure}[h]
\centerline{
\psfig{figure=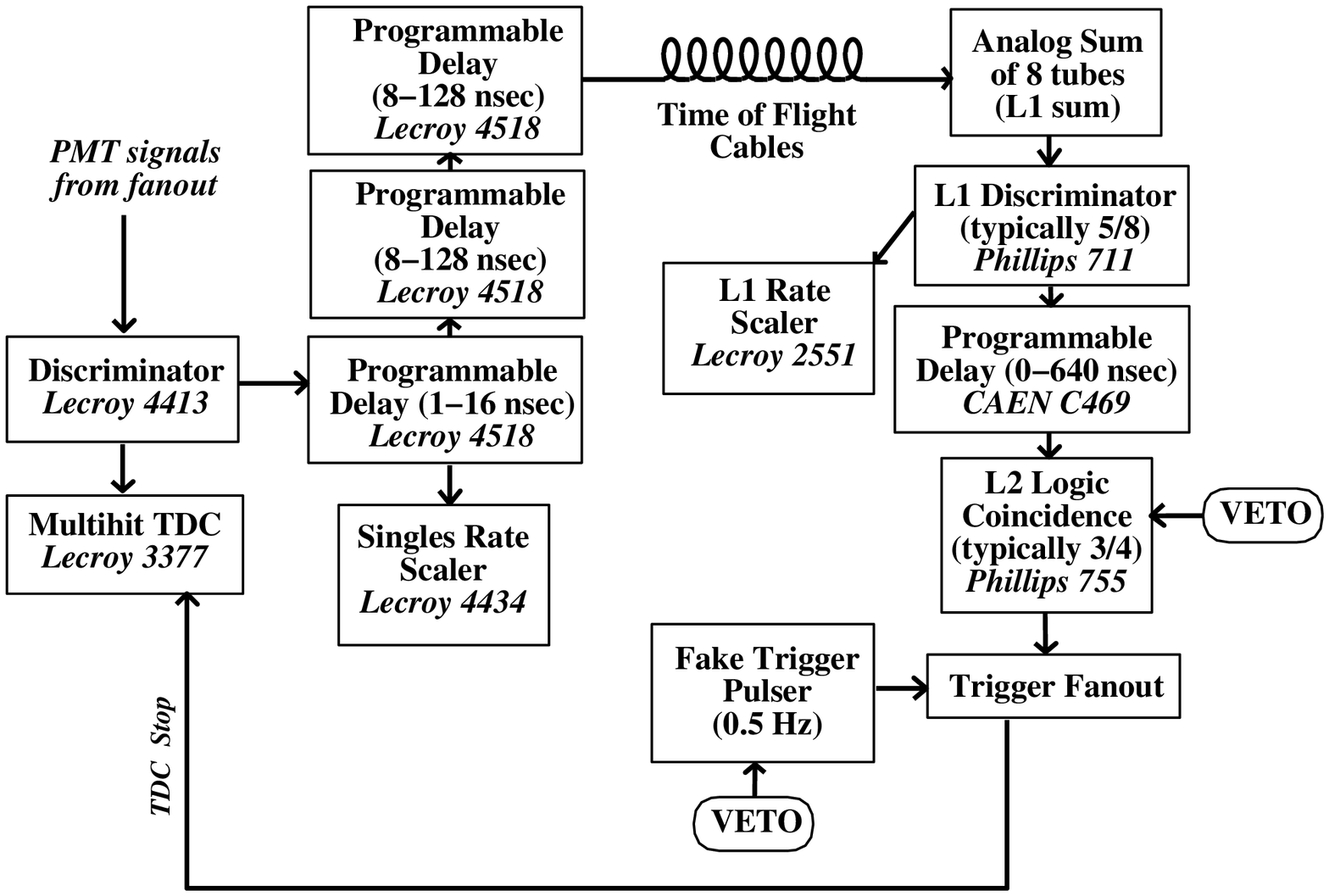,height=3.0in}}
\vspace*{0.5cm}
\caption{STACEE-32 Trigger System with associated hardware. 
Only one channel is shown. 
Pulses from the phototubes are discriminated, delayed and combined 
in groups of 8. 
The sums are then subjected to a multiplicity requirement to 
form a level 1 (L1) or subcluster trigger. 
L1 triggers from the 4 subclusters are delayed and sent to multiplicity 
logic where the level 2 (L2) or master trigger is formed.
Not shown are ECL-NIM converter modules and a pulse reshaper which
follows the time-of-flight delay cables.} 

\label{trigger}
\end{figure}

A schematic of the STACEE-32 trigger electronics is shown in 
figure~\ref{trigger}.

\begin{figure}[h]
\centerline{
\psfig{figure=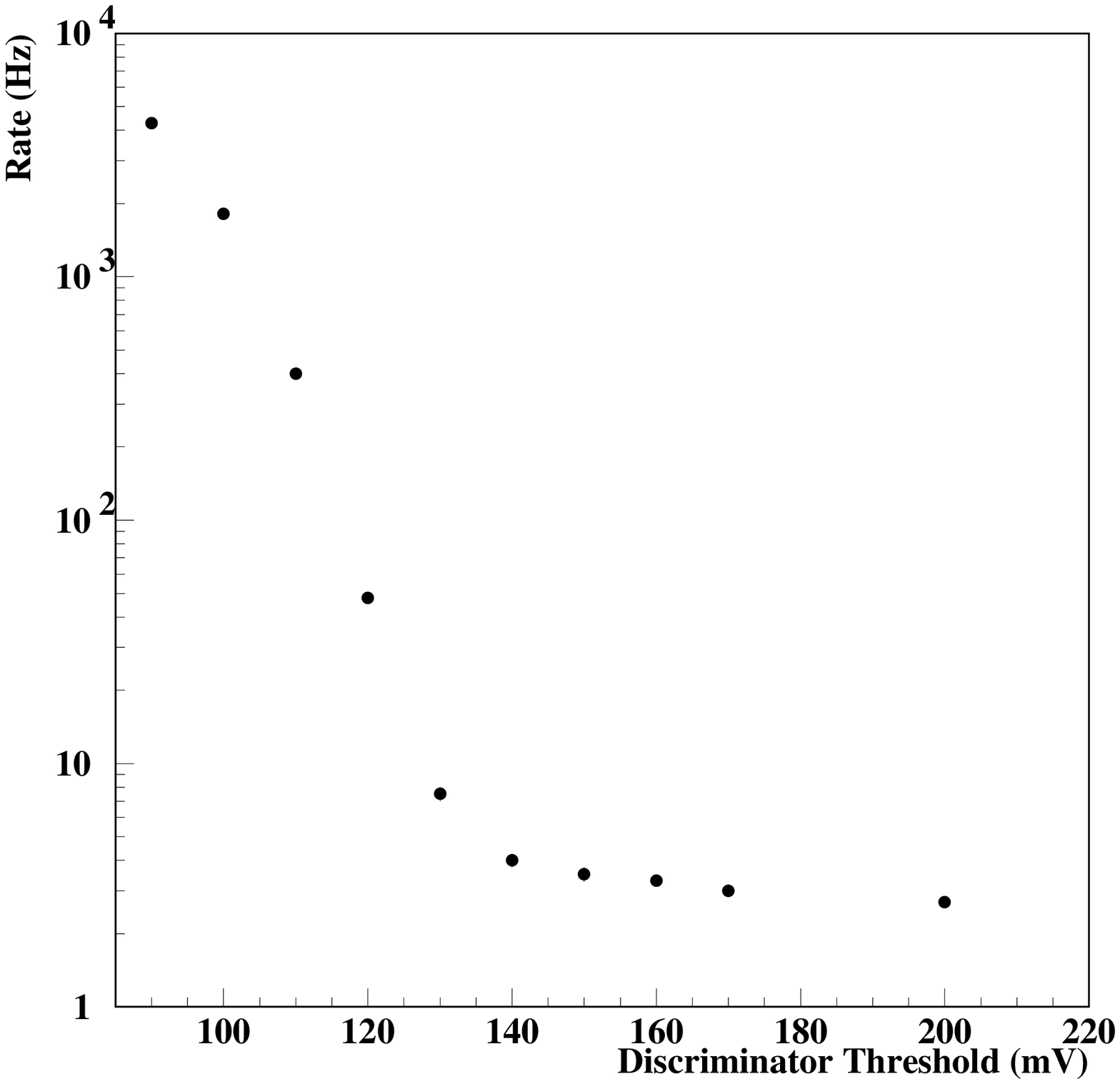,height=3.0in}}
\vspace*{0.5cm}
\caption{Rate $vs$ threshold curve for STACEE-32.  Plotted here is the 
L2 rate as a function of the threshold
on the discriminators receiving the PMT pulses.
Below the ``break point'' at 130 mV, the rate climbs exponentially due to accidentals from 
night sky background. 
Above this value the curve is flatter and is due primarily to 
Cherenkov light from air showers.
The mean amplitude of a pulse due to a single photoelectron is $\sim$20 mV
so the break point occurs at approximately 6 photoelectrons.}
\label{rate_thresh}
\end{figure}

All analog signals from the cameras were discriminated in 16 channel 
discriminators (LeCroy 4413) operating with a common threshold.
This threshold was set according to a rate $vs$ threshold curve like the one shown in 
figure~\ref{rate_thresh}.
In this plot one sees a ``break point'', defined as the
threshold below which the rate climbs exponentially.
The location of the break 
point depended on the individual channel rates, the widths
of the discriminator pulses and the number of channels required to form a trigger.
At thresholds below the break point, 
the rate was dominated by accidental coincidences
and at very low values it flattened due to dead-time effects.
At thresholds above the break point the rate decreased slowly with threshold. 
Here, the experiment was triggering largely on 
Cherenkov light from cosmic ray showers. 

In running STACEE-32 we set the threshold 15-20 mV above the break point, 
which means there was very little background from accidental triggers. 

The accidental rate $R_n$ for a trigger requiring $n$ 
channels to fire, from a group of $m$, each firing at rate $R$ and with 
pulse widths of $\tau$ is given by:

\begin{equation} \label{eq.accidental_rate}
R_{n} = n {m \choose n} R (R \tau)^{n-1}~.
\end{equation}

This formula shows that the accidental rate varies with R$\tau$
to a large power.
To reduce the contribution from 
accidentals, one can either lower R, by raising the 
discriminator threshold or use a narrower pulse width, thus making
$\tau$ smaller. 
Since the discriminator threshold is directly related to the 
minimum number of photoelectrons detected
(and thereby the energy threshold of the detector), it
is advantageous to reduce $\tau$ as much as possible.
With STACEE-32 it was not possible to make $\tau$ less than 
approximately 20 ns.
This width is much longer than the physical limit given by the intrinsic
width of the Cherenkov wavefront (about 4 ns) so there is 
obvious room for improvement. The final version of STACEE will have
new delay units that do not impose this limitation and we will be able to 
run at a lower threshold.

STACEE-32 had a two-level trigger. 
Two levels were necessary because of the limited ranges
of the programmable delays described below. 

The 32 heliostats were divided into four local subclusters of
eight heliostats each, as shown in figure~\ref{h-layout}.
The discriminator outputs from the eight channels in 
each subcluster were routed through delays programmed
to bring in-time hits into coincidence.
The in-time signals were added in a 
passive summing circuit. 
The output of this circuit was sent to a discriminator with a 
threshold set corresponding to 
a certain multiplicity, for example 5/8. 
This multiplicity requirement 
constituted the ``Level 1'' (L1) or subcluster trigger.

The programmable delays were 
accomplished by commercial modules (LeCroy 4518).
These delays compensated for the changing timing of
the wave-front 
caused by the apparent motion of the source across the sky. 
They were supplemented by fixed delays, made of low loss RG213 cables,
which accounted for the different distances of heliostats from the 
tower. 
Signals emerging from the delays were re-shaped by a discriminator with 
a 12 ns output width. 
Typical PMT rates were 1-5 MHz and typical L1 rates were 1-10 kHz.

Each L1 trigger was routed through a second delay unit (CAEN C469) which
adjusted for the time differences among the four subclusters.
These differences were order 200 ns,
larger than the intracluster delays owing to 
the longer distances involved.

The LeCroy 4518 delays used tapped delay line technology so they were
effectively deadtimeless.
The PMT discriminators contributed an effective deadtime 
that was rate dependent, typically less than 
5\% per channel.
Individual channel deadtime was studied for on-source and off-source
data for each source tracked, to ensure that its effect canceled in the 
on-off subtraction.
The CAEN units used digital counting techniques to set the delay
so were dead during each delay period.

The Level 2 (L2) trigger was formed by requiring a minimum
multiplicity from the delayed L1 triggers, typically 3/4.
This multiplicity requirement was implemented using 
a standard logic unit (Phillips Scientific 755)
with input pulses of 12 ns width.
Trigger formation was inhibited whenever an external veto signal was 
asserted.

A by-product of the two level trigger (the L2 requirement) is that 
the light pool is required to be more spread out over the entire
detector. 
This is a feature expected of showers due to gamma-rays.
The multiplicity requirements for the L1 and L2 triggers were 
selected to give the lowest possible energy threshold consistent
with good gamma-ray sensitivity over the entire array. 
A figure of merit which was the Crab gamma-ray rate divided
by the square root of the rate for charged cosmic rays was used
in simulations to quantify the studies~\cite{oser00a}.

L2 triggers were combined with fake trigger signals generated at a 
rate of 0.5 Hz.
Fake triggers were used for determining individual channel
rates and pedestal values.
Whenever either type of trigger occurred (L2 or fake), a common stop signal 
was sent to the TDC, a GPS clock time was latched and 
ADC integration gates were sent to the ADCs (see next section).
Event read-out was then initiated.
During readout, a veto was asserted to prevent occurrence of additional
triggers. 
The veto was cleared by the data acquisition program (DAQ) at the conclusion 
of readout.
The typical readout time was 40 ms per event.

The STACEE-32 delay system had sufficient range to trigger on Cherenkov
showers coming from any region of the sky within $45^o$ of zenith.
Individual delay settings could be controlled with nominal precision
of 1 ns for the L1 delays and 2.5 ns for the L2 delays.
To ensure precise timing, every channel was calibrated with test pulses
and the calibrations were employed in a lookup table.
The delays were updated every few seconds as STACEE-32 tracked sources 
across the sky.

Accidental rates due to random night sky hits in the PMTs could be 
directly measured by using random settings for delays, thereby  
imposing unphysical coincidence requirements. 
Under normal operating conditions the accidental rate was 
5-10 triggers per hour.

\subsubsection{Charge Measurement}

\begin{figure}[h]
\centerline{
\psfig{figure=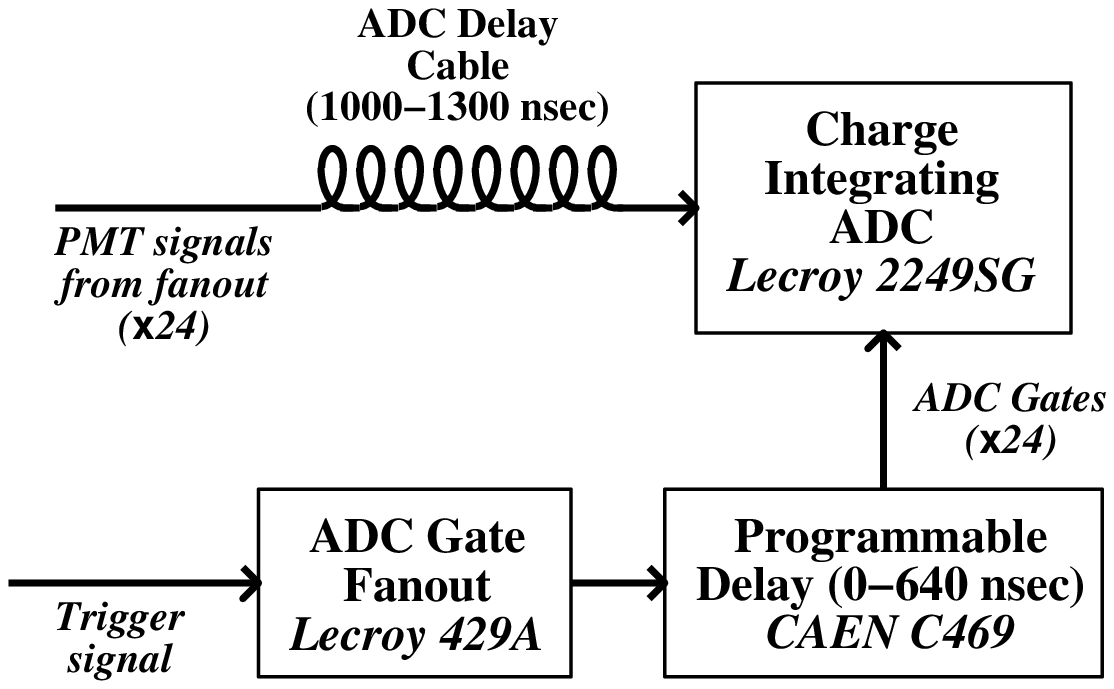,height=2.5in}}
\vspace*{0.5cm}
\caption{STACEE-32 ADC System. 
Analog signals from the linear fanout are routed through cables
to delay their arrival time at the ADC input until after a trigger
has been formed. A dedicated  system  delays integration
gates for each channel, putting each gate in time with the incoming PMT
pulses from the delay cable. 
Twenty-four of the thirty-two channels were instrumented with ADCs.}

\label{ADC}
\end{figure}

Analog pulses from the PMTs were digitized using Wilkinson-type
integrating ADCs (LeCroy 2249SG) which are 11 bit devices with 
0.25 pC/count resolution. 
There are 12 channels per double width CAMAC module and each channel 
has its own gate.
This latter feature was required because of the different and changing
times between signals in different PMTs.
Signals routed to the ADC inputs passed through long (1000 - 1300 ns)
low-loss RG213 cables which delayed the signals while the trigger was formed,
a task that took of the order of a microsecond.
The ADC gates were made by fanning out a digital pulse from the trigger and 
delaying each copy by the appropriate amount in a CAEN C469 programmable
delay module.
The gates arrived at the ADC module 5 ns before the pulse and, as with 
the trigger, the delays were updated as the source was tracked across 
the sky.
This scheme is depicted in figure~\ref{ADC}.
Due to budget constraints, only 24 of the 32 channels in the detector
were instrumented with ADCs.

Routing the PMT signals through long cables had the effect 
of distorting the pulses.
The main effect was to make each pulse longer and smaller in amplitude.
(This is discussed more fully in the section on ADC modeling.)
To accommodate the lengthened pulse, we set the gate width to 37 ns.
This long gate allowed a significant amount of night sky noise to 
be added to the signal.
Assuming a typical single photoelectron counting rate of 1.5~GHz on each 
PMT, an average of 56 night sky background photoelectrons arrived during the
integration gate.
Since the PMTs were AC-coupled, the error on the pulse measurement was 
the square root of this or about 8 photoelectrons.
This meant that the charge resolution of the ADC system was poor near the 
threshold of about 6 photoelectrons.
The utility of the ADC system was largely in allowing time slewing 
corrections to be made to the data.

As part of the upgrade to the final detector 
we are equipping all channels with 
1 GS/s digitizers (FADCs). The digitizers 
will obviate the need for delay cables
since the PMT pulses can be retrieved from the appropriate location 
in the digitizer memory.

\subsection{Laser Calibration System}

STACEE-32 was equipped with a laser calibration system comprising 
a $100~\mu$J nitrogen laser and dye cell feeding a network of optical 
fibres through a system of adjustable neutral density filters. 
The fibres  delivered light to the PMT's by exciting small wavelengthshifter
plates attached to the center of the secondary mirrors. 
The intensity of each laser shot was measured independently using 
4 PIN photodiodes. 
A very similar system is described more fully in ~\cite{hanna00}.

\begin{figure}[h]
\centerline{
\psfig{figure=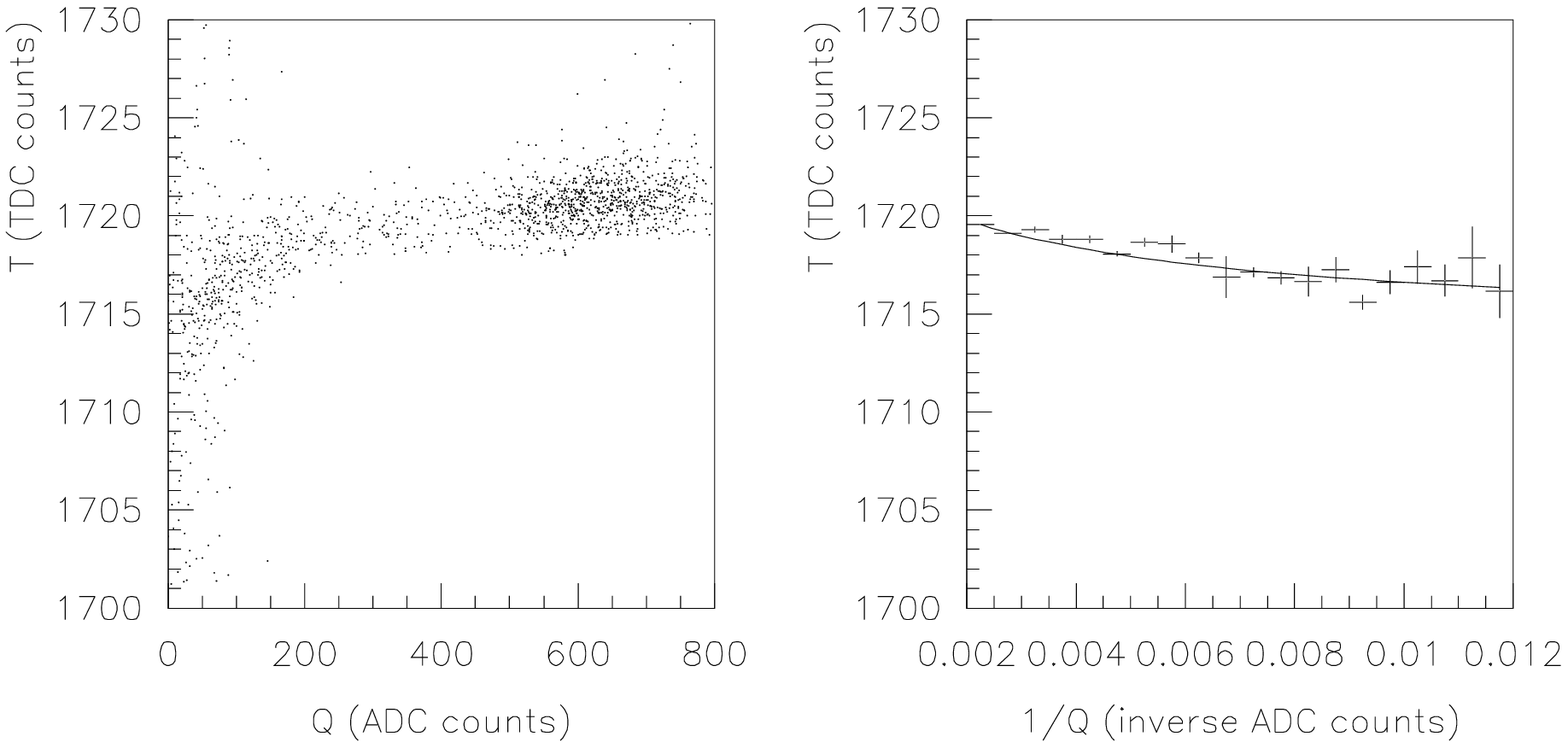,height=4.5in}}
\vspace*{-5.0cm}
\caption{Pulse Slewing Measured with the Laser System.
Large amplitude pulses cross a fixed voltage threshold earlier than smaller
amplitude pulses, resulting in a systematic dependence of the measured
time on the amplitude of the pulse.
This effect can be corrected by adjusting the time according to 
the measured charge of the pulse. 
The effect is illustrated in the left hand plot 
where TDC time (1 count = 0.5 ns) is plotted
against the corresponding ADC value (1 count = 0.25 pC) for a single PMT.
Larger TDC counts correspond
to earlier times.
The effect can be corrected for using a function ($t = a + b/(Q+c)$)
shown fit to the data in the right hand plot where the average TDC value is plotted against the inverse of the charge.}

\label{slewing}
\end{figure}

The system was used for measuring the relative time differences 
between PMT channels.
In the final detector, it will also be used for monitoring gain stability of the PMTs.
A key application in STACEE-32 was the parameterization of time-slewing
effects. 
Time slewing is the phenomenon whereby pulses of different amplitude
exceed the discriminator threshold at systematically different times.
This effect can be studied by sending a series of laser pulses covering a 
range of intensities to the PMTs.
By plotting the measured time for a given channel against the corresponding
charge, as in figure~\ref{slewing}, one can 
fit a parameterization of the slewing effect to the data.
(The parameterization chosen is an empirical function which describes the
data; it has no physical significance.)
Fits were performed for all 24 ADC-equipped channels and the parameters
obtained were used in data analysis.
The correction can be up to about 3 ns, which is large relative to 
the desired timing resolution of  1 ns.

The timing resolution was estimated by examining distributions of residuals
to shower fits. 
It was better than 1 ns for all channels and did not depend
on pointing angle.
It was stable over time.  
 
\subsection{Miscellaneous Electronics}

STACEE-32 had a GPS clock (Truetime GPS II) which was used
to provide a time-stamp, accurate to $1~\mu$s,
for all recorded events. 
These time-stamps were necessary for pulsed emission searches.

As shown in figure~\ref{electronics}, counters were used to measure
the fraction of time that the experiment was live.
Whenever a trigger occurred, a flip-flop was set which asserted a veto signal 
that inhibited further triggers. 
This veto was required to prevent additional triggers from interfering 
with the readout, and it was cleared by the data acquisition computer
once the readout was complete. 
A 10 kHz clock was sent to 
two scaler channels to measure the deadtime.
One scaler counted a vetoed copy of the clock and the other counted
an unvetoed copy.
The ratio of the vetoed scaler count to the unvetoed scaler count 
gave the livetime fraction of the experiment.
Additional scalers were used to monitor the rates of all the
phototubes as well as the L1 subcluster triggers.

The L1 deadtime fractions during normal running were negligable.
The L2 combined L1 rates that were typically of order 10 kHz 
and used delays of order 200 ns.
Thus the deadtime introduced by the non-pipelined CAEN delays was not 
important.
The overall deadtime was dominated by the readout time and varied between 
8\% and 12\% depending on the trigger rate.

\subsection{Data Acquisition}

The data acquisition (DAQ) system for STACEE-32 was based on a Silicon 
Graphics Indy computer which was coupled to two CAMAC crates
via two Hytec Ethernet crate controllers. 
Data were read out after each event trigger and stored on a local disk.
At intervals the data were written to DLT tapes for archiving and 
off-site analysis.

Processes running on the DAQ computer included run control (starting,
stopping and clearing of vetoes), 
monitoring of PMT currents and rates, generation 
of alarms, and readout of the ADC's, TDC's, scalers and the GPS clock.
The DAQ system also calculated and set trigger delays using the CAMAC
delay modules.
It was also used to set discriminator thresholds and high voltage values.
 
\clearpage

\section{Performance}

In STACEE-32 the experimental setup 
involved the atmosphere as part of the detector
so its behavior cannot be verified by a test beam. 
Instead, it is modelled by Monte Carlo
programs and 
certain aspects of these programs can be checked
with data.
In this section we report on results of some of these calculations
and tests.

The modeling of STACEE-32 can be divided  into three parts.
They are:
\begin{itemize}

\item{Simulation of extensive air-showers}

\item{Simulation of the optical throughput of STACEE-32}

\item{Simulation of the electronics}

\end{itemize}

In the following subsections we deal with each part in turn.

\subsection{Extensive Air-Shower Simulation}

\subsubsection{The MOCCA Monte Carlo Program}

The design and understanding of STACEE-32 was aided by the MOCCA 
air shower simulation package \cite{hillas85}.
MOCCA simulates the entire development of an extensive air shower,
starting with the first interaction of the primary particle in the 
upper atmosphere and following all generated secondary particles 
until they reach the ground or their energy falls below the point
where they no longer contribute to shower development. 
Note that at the energies relevant to STACEE-32 (E $< 10^{12}$~eV)
we are able to follow all particles since the multiplicities are small enough.

With user-specified
energy, incident direction and particle type of the initial primary,
MOCCA's output consists of wavelengths, coordinates and directions
of Cherenkov photons, along with their time of impact at ground 
level. 
The intervening processes accounted for by the program include 
ionization, bremsstrahlung 
and pair production, as well as effects such as deflection by 
the geomagnetic field and Coulomb scattering in the atmosphere.

Our chief goal for STACEE-32 
was the detection of gamma rays so the understanding
of the detector's response to electromagnetic showers is important.
The precise response to hadrons is less important since it is largely needed 
for understanding backgrounds from charged cosmic rays.
MOCCA, like all shower Monte Carlo programs,
has little difficulty with electromagnetic showers since they are 
well described by QED calculations.
Uncertainties arise with hadronic showers, 
which cannot be calculated from first
principles.
Even these uncertainties 
are not expected to be very large since, for collisions in the 
STACEE energy range, most cross sections have been measured at accelerators
or can be reliably extrapolated from lower energies.
MOCCA uses simplified descriptions of these cross sections
to model hadronic interactions.

There are other programs that simulate air showers. 
Another leading package is CORSIKA \cite{heck}, developed 
by the KASKADE collaboration.
CORSIKA makes use of packages such as EGS4 \cite{nelson85} and 
GHEISHA \cite{fesefeldt}, which are widely used in nuclear and particle 
physics.
The more detailed particle interaction models in CORSIKA tend to make it
slower than MOCCA.

To check the robustness of MOCCA predictions of key observables
such as total yield of Cherenkov photons we have compared it with 
CORSIKA. The photon densities from the two simulators agree at the
$ \sim 6\% $ level for primary gamma-ray energies of
50 GeV and 100 GeV and for zenith angles from $0^o$ to $45^o$.
This agreement 
is reassuring but not surprising since an electromagnetic shower
is a well understood and well modelled phenomenon.

The agreement for hadronic showers is not as good, reflecting the 
greater uncertainties inherent in modeling hadronic interactions.
The total photon densities from MOCCA agree with those from CORSIKA
to within $15\%$.

\subsubsection{Atmospheric Modeling}

An important input to the calculations is the assumed atmospheric profile.
The development of the shower depends on the density profile of the 
atmosphere.
The rate and angle of production of Cherenkov photons
by particles in the shower depend on the local refractive index.
This index can be computed from the local temperature and pressure.
In addition, the attenuation of Cherenkov light due to Rayleigh and Mie 
scattering and absorption by oxygen allotropes are important
effects.

The calculations for STACEE-32 used a profile in which the atmosphere was
divided into five zones. 
The density profile followed an exponential distribution with a 
characteristic scale height in each zone. 
The model parameters were determined from Linsley's parameterization 
of the U.S. standard atmosphere \cite{linsley}.

Rayleigh scattering is the most important mechanism for attenuating the
Cherenkov light of interest to STACEE-32.
It scatters photons outside the field of view of the detector with 
negligible chance of scattering back into the acceptance.
The $(1/e)$ absorption length for this process is given by:

\begin{equation}
L = \frac{3}{32\pi} \frac{N \lambda^{4}}{(n-1)^2}~,
\end{equation}

\noindent
where $N$ is the number density of air molecules, atoms and ions (but not 
electrons), $\lambda$ is the photon wavelength, and $n$ is the index of
refraction.

\begin{figure}[h]
\centerline{
\psfig{figure=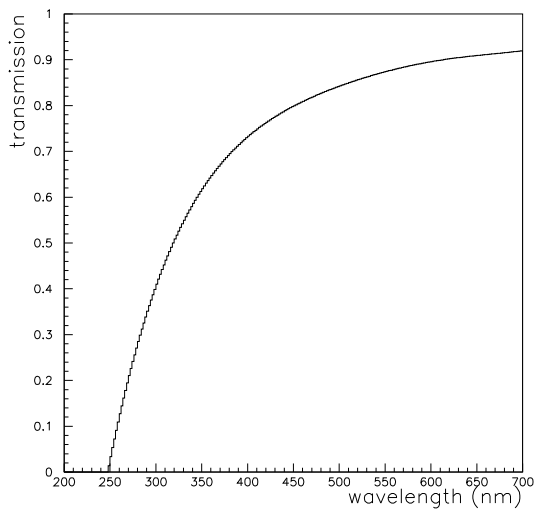,height=3.0in}}
\vspace*{0.5cm}
\caption{Transmission probability $vs$ wavelength 
for the model atmosphere described in the text. 
Plotted is the probability for a Cherenkov 
photon produced in a MOCCA-simulated
air shower to reach the detector altitude.
Different atmospheric models give the same transmission curves
to within $\pm 10\%$.} 

\label{trans}
\end{figure}

For vertical air showers simulated by MOCCA,
the transmission probability for Cherenkov photons 
as a function of wavelength is shown 
in figure~\ref{trans}.

\subsection{Optical Simulations}

The second part of the STACEE-32 simulation chain traced the optical path of
Cherenkov photons through the detector elements.
For this part, a custom-made ray tracing package called ``Sandfield''
(Sandia Field Simulator) was developed.
Sandfield followed the path of every Cherenkov photon through the 
optical elements (heliostats, secondary mirrors and DTIRCs) onto the 
PMT photocathodes, folding in transfer efficiencies at every stage.
The end result was a list, for each channel, of photo-electrons and 
their times of arrival. These lists were passed to the electronics simulator
for further processing.

In the following, we describe in more detail some features included in 
modeling the different optical elements of STACEE-32.

\subsubsection{Heliostat Modeling}

As described earlier, the STACEE-32 heliostats were compound structures
consisting of 25 individually adjustable facets mounted on a frame which
itself could be oriented in two directions. 
Each facet was a 4-foot square of back-aluminized glass which could 
be deformed to a shape which was, to first order, a paraboloid.
The facets were co-aligned to bring the image of the Sun to a tight 
spot on the central tower. 

Sandfield modelled a heliostat as a collection of 25 parabolic reflectors, 
each with 
a focal length equal to the distance from the heliostat to the tower.
The orientation of each facet with respect to the heliostat frame was 
individually adustable within the program. These orientations were set to 
their correct values (those that would produce the smallest possible
sunspot) then given a random Gaussian error in heading.
This error was expressed in terms of a parameter called $\sigma$
which represented the RMS of the linear displacement of the facet image 
on the tower.
The displacement was measured radially from the nominal target position. 
Typically $\sigma \sim 0.5~$m.

\begin{figure}[h]
\centerline{
\psfig{figure=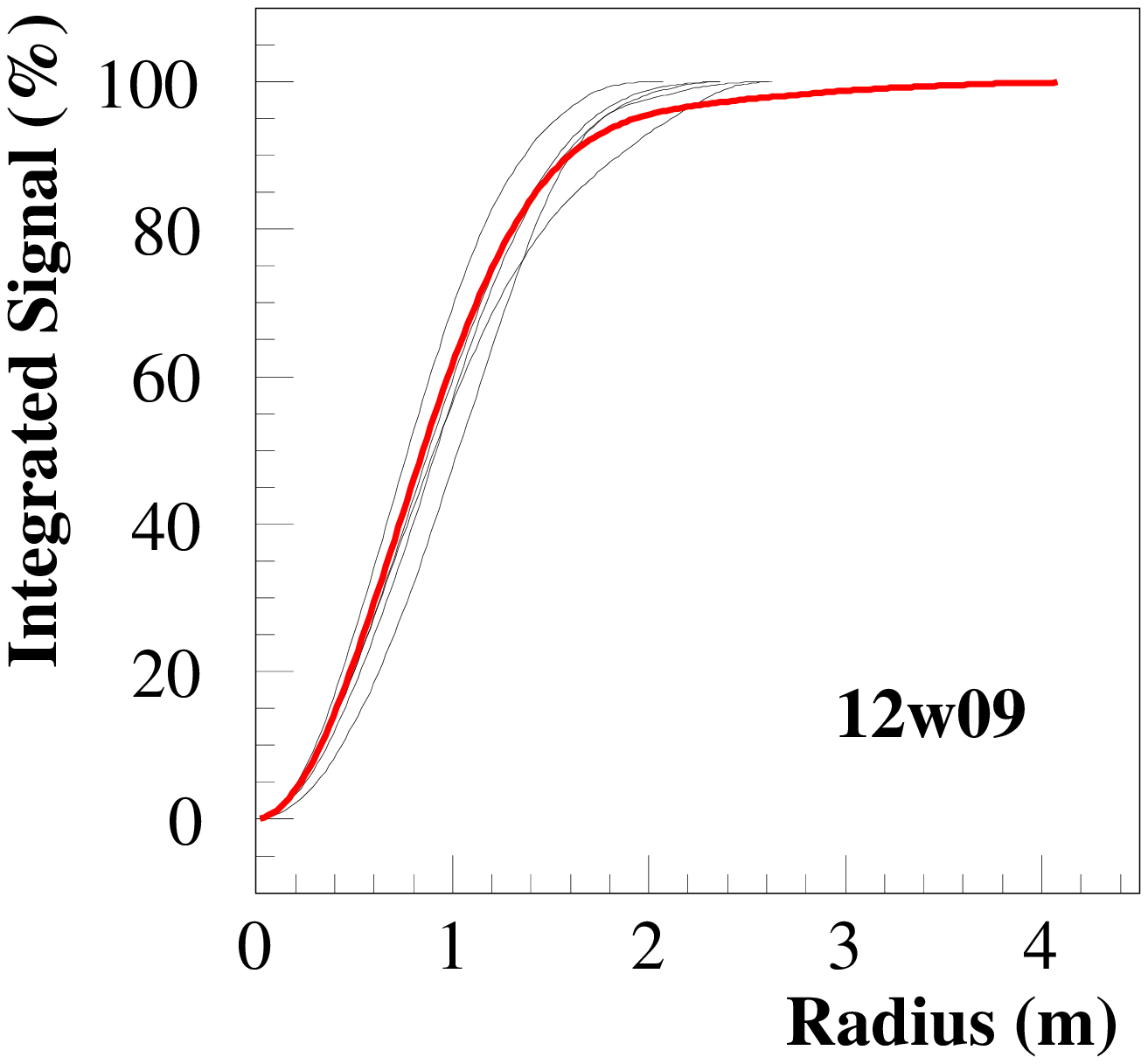,height=4.0in}}
\vspace*{0.5cm}
\caption{Simulated and Measured Sunspot Curves for a STACEE-32 heliostat
(12w09).
The integral fraction of total sunspot light contained within a circle on the 
tower is plotted as a function of the circle's radius.
The thick solid curve is the measured sunspot contour, 
obtained from an image made with a CCD camera. 
The other curves are from simulated sunspots. 
Each of the 5 simulated curves was made with the same set of parameters
except for the orientations of the individual facets, which were randomized.
} 

\label{sunspot}
\end{figure}

The value of 
$\sigma$ for each heliostat was determined by comparing simulated sunspots,
made with different values of $\sigma$, with the real sunspot, as measured
using a CCD camera.
In figure~\ref{sunspot},
simulated and measured sunspot curves for a sample heliostat
are shown.
Plotted are the normalized integrals of light as a function of
radius, $ie$ fractions of total light contained within a given
radius. 
The simulations have been made with a value of 
$\sigma$ that 
provides the best fit to the measured data.
Typically a $1~$m radius secondary captured about $60\%$ of the light.

\subsubsection{Secondary Mirror and Camera Modeling}

Cherenkov photons reflecting from the heliostats were traced to 
the tower where they encountered the secondary mirrors.
The photons were then directed towards the cameras. 
Some of the photons from heliostats near the tower were occulted by the 
camera structures and did not survive to hit the secondaries; this
effect was accounted for in the simulation. 
The secondaries themselves were modelled as single-piece spherical 
mirrors with reflectance as shown in figure~\ref{reflect}
and were oriented as determined during a survey and alignment procedure
during installation. 
The faithful treatment of the spherical optics automatically produced
the spherical and off-axis aberrations which were important features
of this setup.

\subsubsection{DTIRC and Photocathode Modeling}

Sandfield included ray tracing routines which simulated the DTIRCs
and the silicone rubber disks which coupled them to the PMTs.
Dielectric reflection and refraction effects were included at all
medium interfaces.
Absorption inside the DTIRC and attenuation 
effects in the silicone were also included.

The last step in the optical modeling of STACEE-32 was at the 
PMT photocathode. 
Here the photons were converted to electrons with a probability
equal to the quantum efficiency of the PMT. 
This quantity depended on the wavelength of the incident photon 
and composition of the photocathode. 
Additionally, it depended on the angle of incidence and point 
of impact of the photon on the photocathode.
The angular dependence has been modelled and verified experimentally
by the CELESTE collaboration \cite{denaurois00} who use the same PMTs
as STACEE-32. 
Their model was incorporated into Sandfield; the effect of the 
angular dependence was to enhance the reponse of the system at angles
near to the geometric cut-off, 
where photons were impacting  the photocathode at very oblique
angles.
The enhancement can be seen in figure~\ref{enhance}.

\subsection{Electronics Simulation}

STACEE-32 PMTs were bombarded with a high flux of photons either 
from air showers or from night sky background. 
The elevated rates caused pile-up effects which needed to be understood
quantitatively, so a detailed electronics simulation package was 
essential. 
In this section we describe HERMES 
(Helpful Electronics Reconstruction and Measurement Endcap Simulation),
the package developed for STACEE-32.

\subsubsection{Pulse Library and Construction of Waveforms}

HERMES simulated the behavior of the STACEE electronics starting from 
the analog pulses at the output of the PMTs.
It made use of a library of 2200 single photoelectron pulses
which were obtained by recording the memory buffer of a digital oscilloscope
(Tektronix 640A - 2.0 GS/s sample rate) 
responding to single photoelectron pulses from a 
STACEE PMT, set to a gain of $1.5 \times 10^5$.

Simulated PMT pulses were constructed from library pulses by 
linear superposition.
The process began by combining photoelectrons from the air shower
(generated by the Sandfield program) with random night sky background 
photoelectrons generated uniformly in time according to Poisson 
statistics.
The rate for night sky photoelectrons was an adjustable parameter,
and was determined for each PMT from the ratio of its photocurrent to 
its gain.
The signal and background combinations were used to generate 
a simulated analog waveform by placing randomly chosen library
pulses at the arrival times of each photoelectron. 
The sum of these components was sampled in 50 ps steps by
interpolating between the coarse grained points to define the pulse amplitude
at each point in time.

\begin{figure}[h]
\centerline{
\psfig{figure=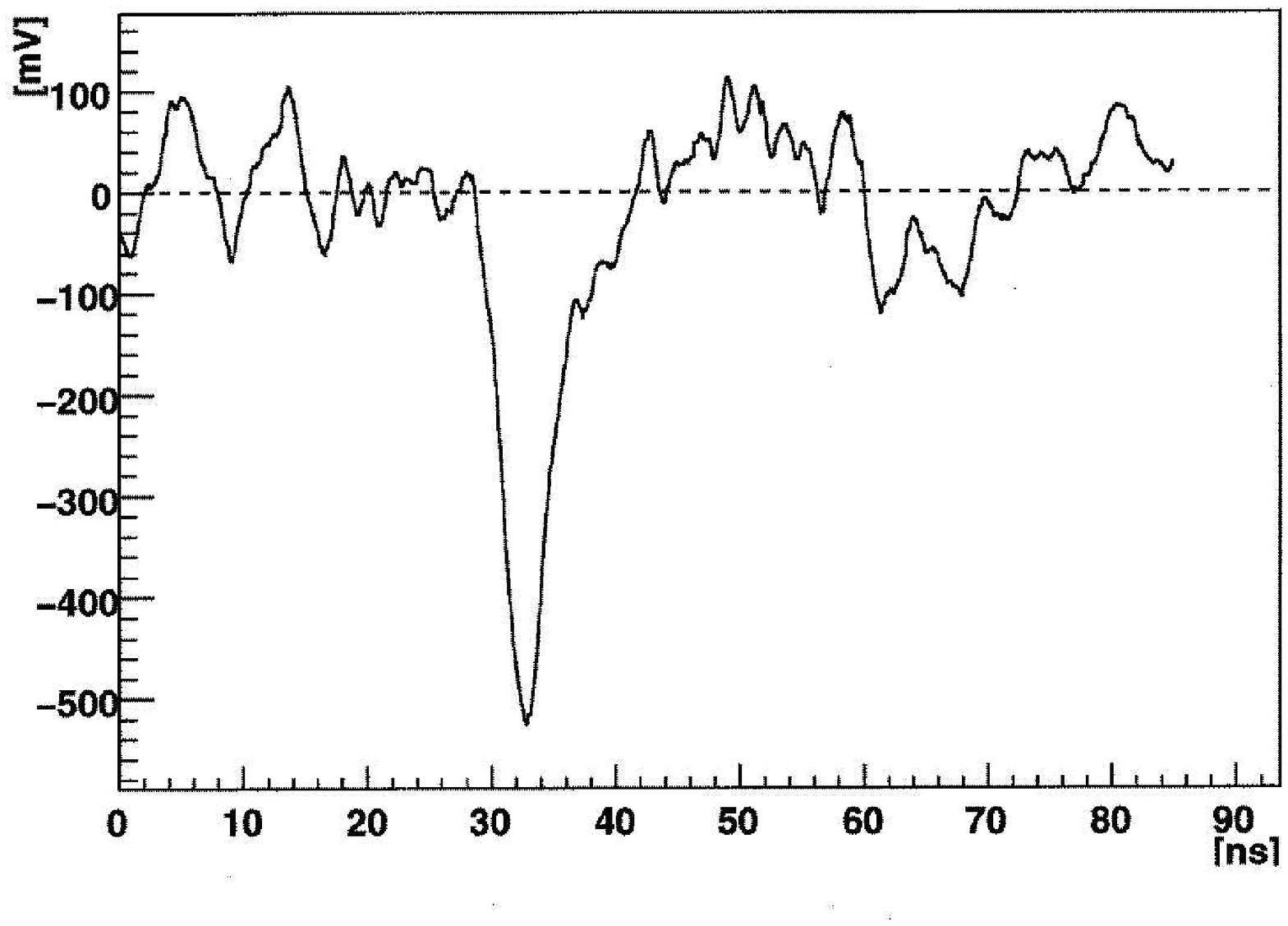,height=3.0in}}
\vspace*{0.5cm}
\caption{A simulated pulse waveform made by adding single photoelectron
waveforms as described in the text. 
The main pulse corresponds to a level of 20 photoelectrons.
} 

\label{fakepulse}
\end{figure}

This process was repeated for all channels resulting in a simulated
event for the experiment. 
Different gains on the channels were simulated by appropriate scaling
of the pulse amplitudes. 
Each waveform was 85 ns long, including a pre-pulse interval of 25 ns; 
a sample is shown in figure~\ref{fakepulse}.

\begin{figure}[h]
\centerline{
\psfig{figure=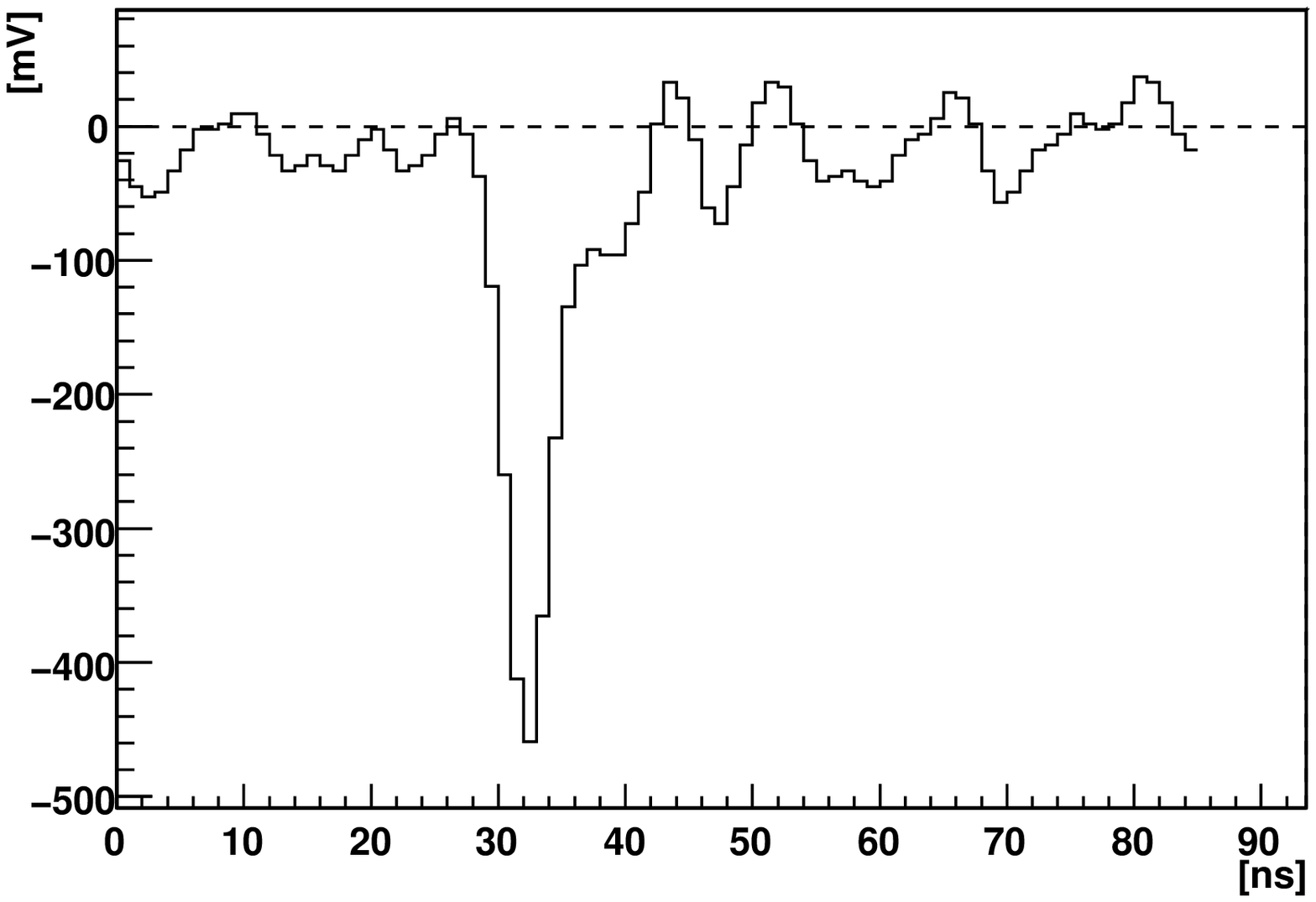,height=3.0in}}
\vspace*{0.5cm}
\caption{A pulse from a STACEE-32 air shower triggered event as captured
by a 2 GS/s digital oscilloscope.
} 

\label{realpulse}
\end{figure}

To show how well HERMES simulated STACEE-32 pulses, we display,
in figure~\ref{realpulse}, a pulse waveform from one channel
of the detector as captured by a 2 GS/s digital oscilloscope while 
running with the standard trigger conditions. 
There is a strong similarity between the waveforms shown in figures
~\ref{fakepulse} and~\ref{realpulse}.

\subsubsection{Discriminator/Trigger Model}

At this point, the simulated data consisted of 32 waveforms.
The waveforms were passed through simulated discriminators which
modelled the behavior of the LeCroy 4413.
Features such as threshold, double pulse resolution, updating behavior,
and hysteresis effects were simulated. 
Deadtime (order 10 ns) after firing was automatically accounted for.

To check the validity of the electronics simulation we compared
certain key quantities. 
For example, the widths and characteristically asymmetric 
shapes of the TDC distributions for real and
simulated data were compared and have been found to be in good
agreement.

A more powerful check is to reproduce the absolute rate $vs$ threshold 
curves for various PMT channels.
The PMT rate at low threshold was due to the random pile-up
of single photoelectrons exceeding
the discriminator threshold.
It was directly related to the photocurrent, both quantities 
depending on the single photoelectron rate, the PMT gain, and the 
average pulse width.
Small variations in any of these quantities produced large
changes in the rate. 
Thus, reproducing actual rate $vs$ threshold curves was a sensitive 
test of the validity of the simulation, particularly the calibration 
of the discriminator threshold level in terms of photoelectron equivalents.
This calibration directly affected the energy threshold calculation.

\begin{figure}[h]
\centerline{
\psfig{figure=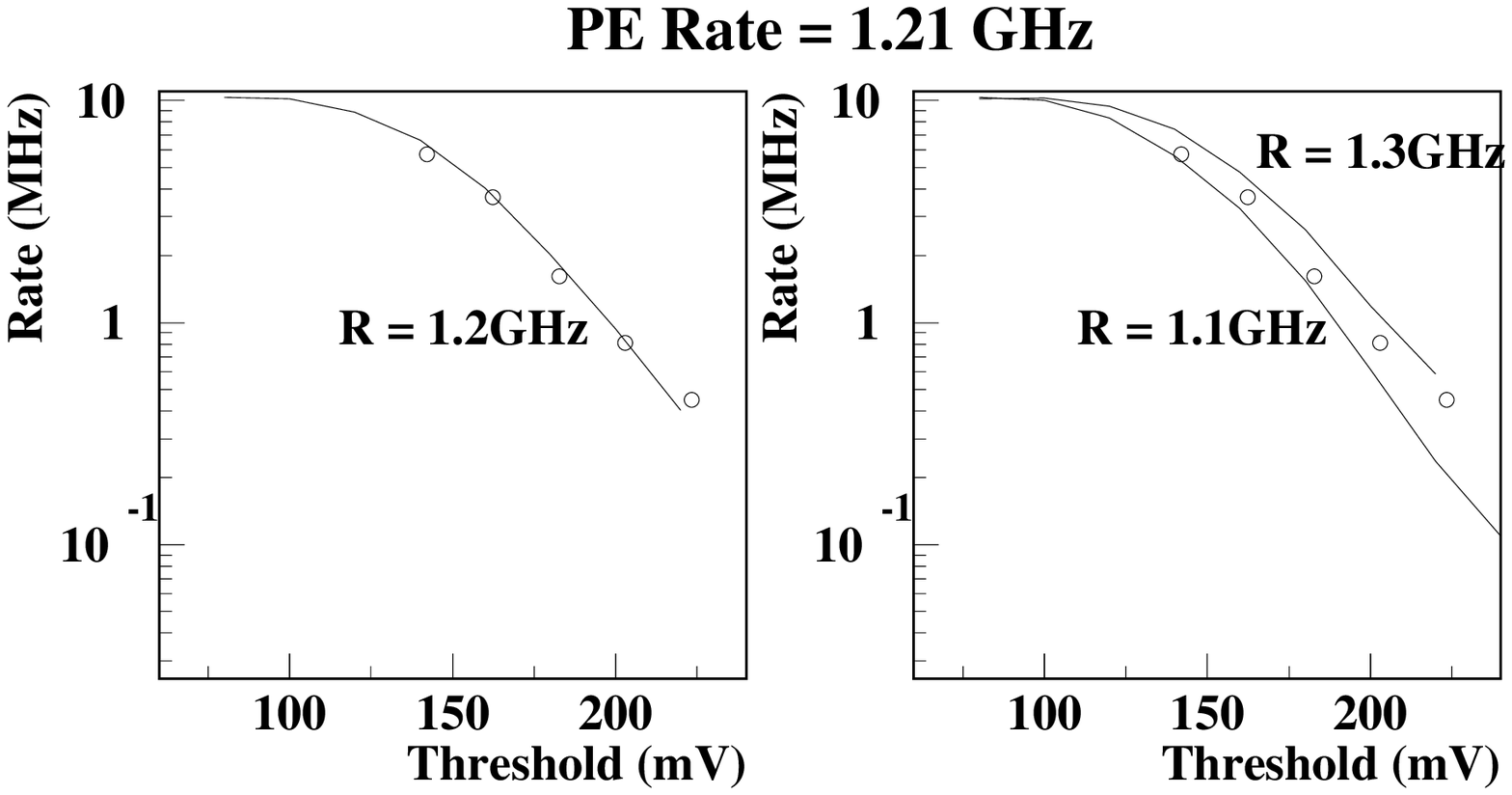,height=3.0in}}
\vspace*{0.5cm}
\caption{Simulated discriminator firing rate $vs$ discriminator
threshold. 
Left: the open circles are rates measured at different thresholds 
for a STACEE-32 PMT operating 
$in~situ$ at an estimated single photoelectron rate of 1.21 GHz.
The curve is a simulation produced for the same gain and single 
photoelectron rate. 
Right: Changing the single photoelectron rate by 
a small amount from the true rate leads to a disagreement between the 
simulation and the data.}
\label{ratecurve}
\end{figure}

A representative simulation is shown in figure~\ref{ratecurve}
where the absolute rate, measured $in~situ$, is plotted against
the discriminator threshold.
The PMT current divided by the calibrated gain gives a single
photoelectron rate of 1.21 GHz.
A simulation curve, assuming this single photoelectron
rate, fits the data remarkably
well whereas curves made with $\pm 10\%$ changes to the single
photoelectron rate fail to describe the data.
(Note that this rate value is peculiar to this channel. 
There are channel-to-channel variations in the exact rate and these
are well described by our simulations.~\cite{oser00a})

\subsubsection{ADC Modeling }

As previously stated, PMT pulses 
were sent through delay cables before being digitized in the STACEE-32 ADCs.
The cables had the effect of considerably lengthening the pulses, leading 
to the requirement of a long (37 ns) ADC gate.
The pulse lengthening was modelled in HERMES using a simple transform
function, $Z$, which had the following effect on a delta function:

\begin{equation}
Z (\delta (t)) = \left\{ \begin{array}{ll}
0 & \mbox{if $t<0$}\\
\lambda e^{-\lambda t} & \mbox{if $t \ge 0$} 
\end{array} \right.
\end{equation}

\noindent
The transform of any arbitrary waveform is found by convolving the 
waveform with this function.

This model was tested using a sample of digitized PMT pulses made with 
an LED flasher before and after they had passed through a long cable. 
The empirically determined values for $\lambda$ varied linearly from 
0.071 ns$^{-1}$ for a 1000 ns cable to 0.048 ns$^{-1}$ for a
1300 ns cable.

The model reproduced the observed amplitude attenuation and the
charge collected within the ADC integration gate.

\subsection{Simulation Results}

The most important application of the simulation programs was the 
determination of STACEE-32's response to gamma ray and cosmic ray
air showers.
To determine flux levels from a source one needs to know the effective
area of the detector as a function of energy.
We used the simulation to estimate STACEE-32's energy response.

\subsubsection{Cosmic Ray Trigger Rate}

A good end-to-end test of the simulation chain is the calculation of 
the rate due to cosmic ray triggers.
The cosmic ray spectrum in the energy range relevant to STACEE-32 is 
well known.
Many cosmic ray runs were taken by 
STACEE-32 so a comparison between calculated and measured cosmic ray rates
was straight-forward.
The cosmic ray data were obtained by collecting showers from the zenith.

A suite of 30,000 proton showers scattered over the spatial, angular 
and energy range of the detector was generated. 
The showers ranged in energy from 100 GeV to 1 TeV, following an
$E^{-2.75}$ differential spectrum, and were uniformly distributed
in solid angle within $2^o$ of zenith. 
Each shower was used 10 times by placing its impact point  
at a different random place  
inside a 500 m diameter circle positioned at 
the center of the heliostat array.

The Cherenkov photons in the simulated showers were propagated
through the Sandfield program and the resultant photoelectrons were
analyzed by HERMES.
Night sky background rates from a zenith run were used to calculate
the correct number of  random
photons to add to the signal photons.

The trigger criteria were applied and 
the ratio of accepted showers to generated showers was the trigger
fraction. 
This fraction
was multiplied by the known flux of cosmic rays (protons and
helium nuclei only - heavier nuclei do not contribute significantly
to the trigger rate) to get the predicted trigger rate.

The rate predicted from the simulation, $3.5~\pm~0.3$~Hz,
was in reasonable agreement with the measured rate.
For a run near the beginning of the 1998-99 campaign, the measured
rate was $2.92~\pm~0.06$~Hz.
This provided a measure of confidence in the validity of the
entire simulation chain.

It is worth noting that the observed cosmic ray rate did not remain
constant throughout the campaign. 
There was a drop from 2.9 Hz in November, 1998 to 2.5 Hz in December.
The rate remained at this level for the rest of the campaign.
It is suspected that bias values 
on some of the heliostats shifted.
These shifts degraded the optical alignment of the detector, 
thereby increasing the energy threshold. 
A system of periodic bias checks has since been implemented to 
prevent this from re-occurring.

\subsubsection{Mean ADC Values}

\begin{figure}[h]
\centerline{
\psfig{figure=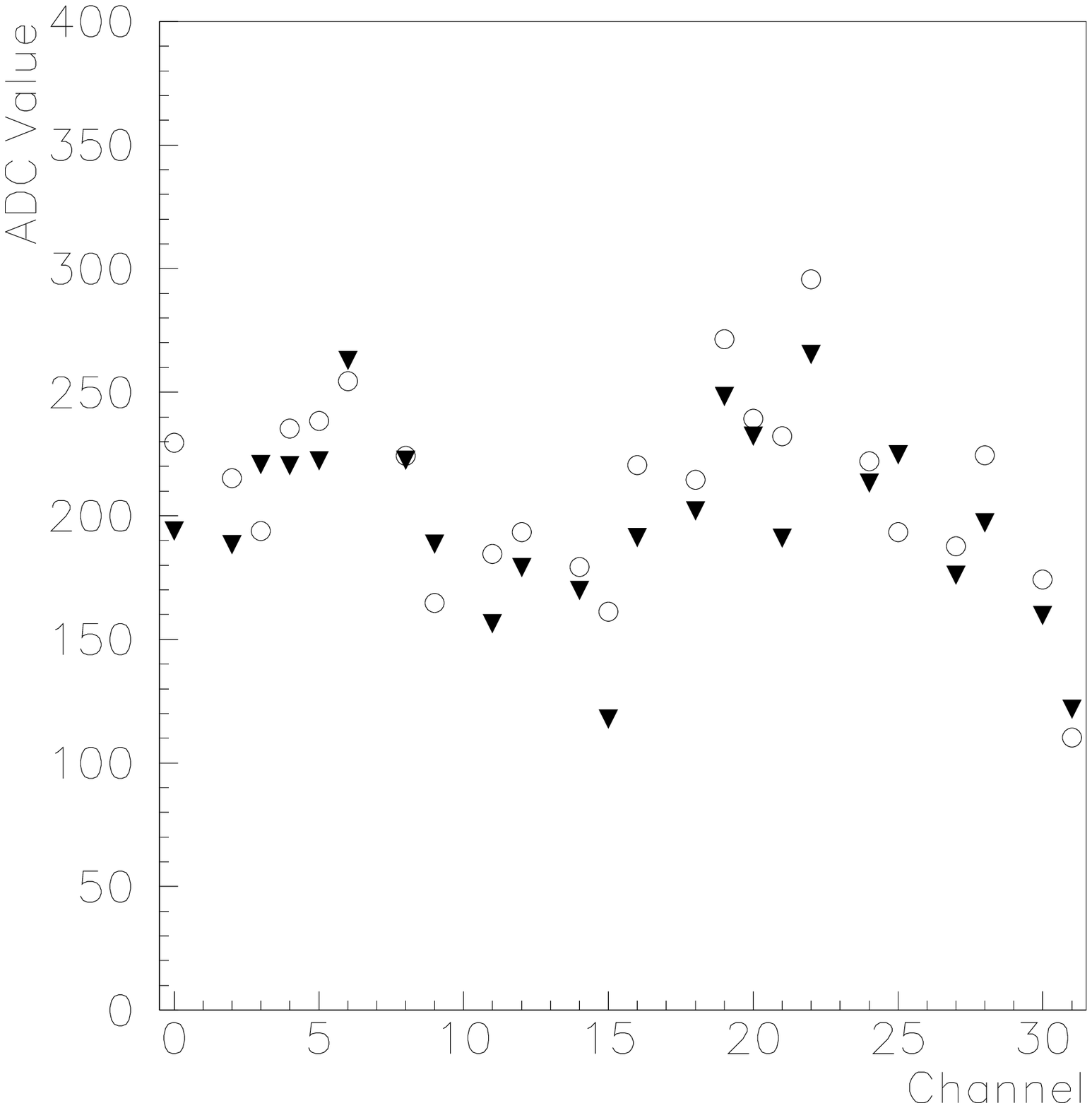,height=3.0in}}
\vspace*{0.5cm}
\caption{Comparison of measured and simulated mean ADC values.
The mean, pedestal subtracted,
ADC values are plotted $vs$ channel number.
Data from cosmic ray triggers are plotted as open circles and 
the results from the simulation are plotted as triangles.
The units are ADC counts where each count corresponds to 0.25~pC.
(Recall that only 24 channels of STACEE 32 were equipped with ADCs
so some channels have no points.)}
\label{adcmean}
\end{figure}

Another end-to-end check of the simulation package was the prediction
of mean ADC values for cosmic ray triggers. 
Using the proton showers described in the previous subsection, the 
mean charge as measured by each ADC was computed and compared with 
that obtained from cosmic ray runs. 
As can be seen in figure ~\ref{adcmean}, there was good agreement 
between simulations and data at the 
$10 \%$ level. 
The channel to channel structure, an artifact of the 
detector geometry, was well modelled.

\subsubsection{Energy Reconstruction}

Air shower energies are usually reconstructed by fitting 
the measured PMT ADC values to a lateral distribution function and
by varying the normalization of the function.
This procedure could be done with data from STACEE-32 
but the resulting resolution 
was expected to be poor, especially near threshold, due to the large
amount of night sky background included in the ADC gates. 
For example, for proton induced showers, a simple sum over ADC channels
gave an energy estimate with a resolution of $\sim 60 \%$.

The situation was better for gamma-ray induced showers coming from
a point source at the center of the detector's field of view.
These showers were smoother and the lateral distribution of
Cherenkov light was more regular than for hadronic showers.
Using a simple sum over ADC channels gave an energy resolution  
of $20 \%$ for 200 GeV gamma rays, averaged over showers whose cores
were distributed uniformly across the heliostat field~\cite{oser00a}.

\subsubsection{Effective Area for Gamma Rays}

A key parameter of any gamma ray detector is its effective 
area. 
This area is, in general, an energy dependent quantity which depends
on the details of the detector elements and the trigger criteria.
We define the effective area as follows:

\begin{equation}
A_{eff}(E) = \int dx~dy~P(x,y,E)~,
\label{eq.eff_area}
\end{equation}

\noindent
where $P(x,y,E)$ is the fraction of gamma rays, the extrapolated 
trajectories of which land at position $(x,y)$ on the array,
which trigger the experiment and are accepted for analysis.

The rate of accepted gamma rays is then:

\begin{equation}
R_{\gamma} = \int_{0}^{\infty} dE~A_{eff}(E)~ \frac{dN(E)}{dE}~,
\label{eq.eff_area2}
\end{equation}

\noindent
where $\frac{dN(E)}{dE}$ is the differential energy spectrum.

To trigger the detector, one needs a minimum density of Cherenkov
photons on the ground. 
Thus, for vertical showers, the effective area rises from zero
at low energy, where the photon density is below trigger threshold, 
to the area of the light pool $(\sim \pi$(100~m)$^2$) at high energy.
For non-vertical showers, the situation is more complicated.
Here, geometric effects cause the light pool to be spread out over
a larger area, thus reducing the photon density. 
Also, the photons must pass through a larger slant depth of atmosphere
which subjects them to increased attenuation. 
These effects cause the energy threshold to increase with zenith 
angle, which introduces an angular dependence into the effective 
area.
In STACEE there are angular effects in the light collection optics
(for example projected areas of heliostats) which further complicate the
dependence of area on angle.

\begin{figure}[h]
\centerline{
\psfig{figure=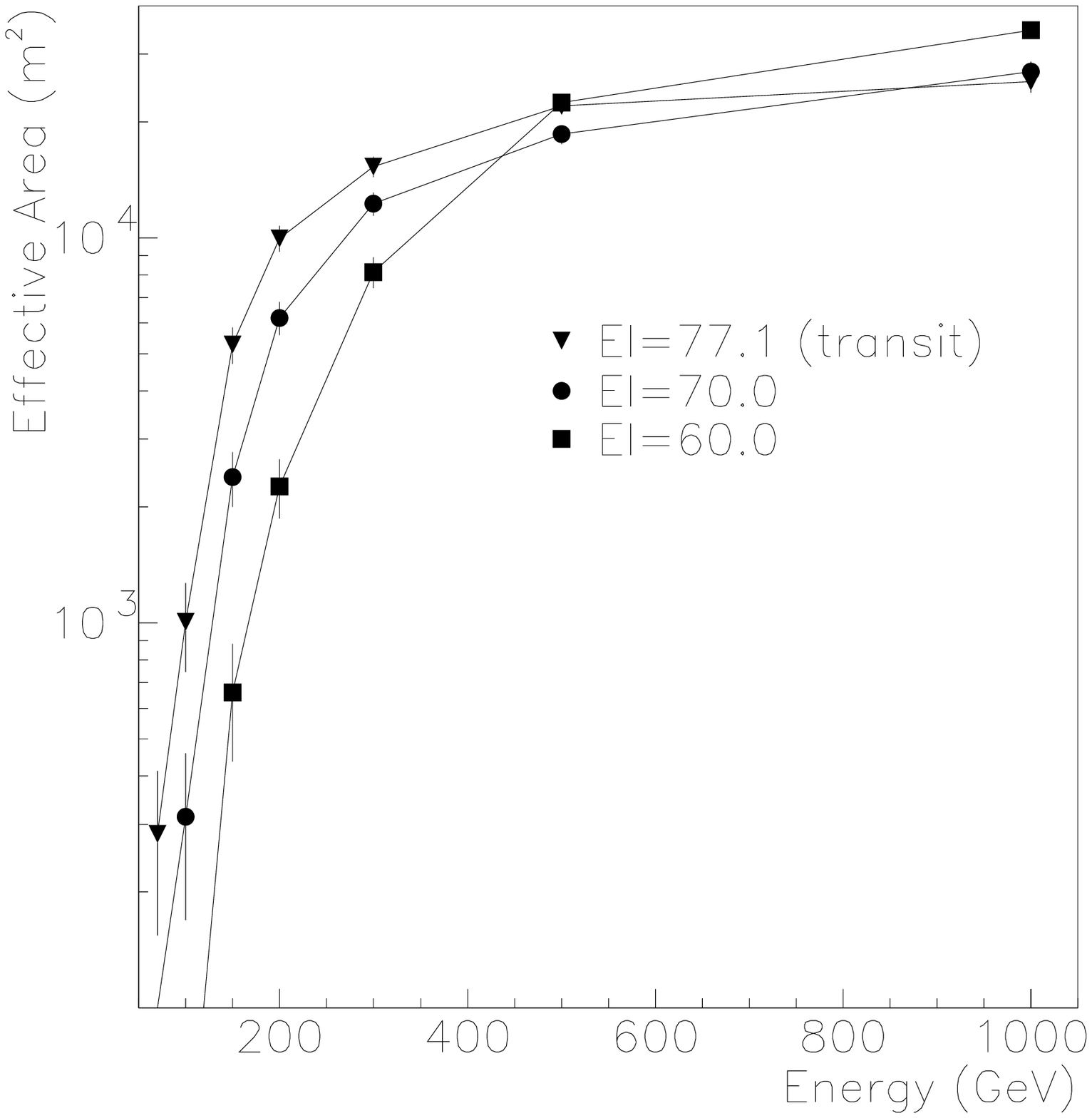,height=3.0in}}
\vspace*{0.5cm}
\caption{Effective area as a function of energy for STACEE-32 for 
three different source elevation angles occurring on the trajectory
of the Crab Nebula.
Each point is based on 4000 shower trials (except for the 1 TeV
points which are based on 2000 trials).
Errors are statistical only.}
\label{eff_area}
\end{figure}

The effective area as a function of energy for a series of angles
was computed using the simulation.
Gamma rays with energies between 50 GeV and  1 TeV were generated
for three different arrival directions corresponding to positions along
the trajectory of the Crab Nebula.
These showers 
were scattered within a 200 m radius circle centered on the heliostat
array, each shower being used for 20 different impact positions.
The standard trigger for the 1998-99 campaign (Level 1 = 5/8,
Level 2 = 3/4, discriminator = $-165$ mV) was applied and the trigger
fraction was determined.
This fraction was multiplied by the area over which showers were 
thrown ($\pi$(200~m)$^2$) to get the effective area. 
The resulting curves are shown in figure ~\ref{eff_area}.
The acceptance is essentially zero below 100 GeV but turns on rapidly
and rises to a plateau value of $\sim$ 28,000~m$^2$ at high energy.
The energy at which the turn-on occurs depends on angle, as expected,
as does the plateau value.
Note that the uncertainties shown in the figure are statistical only. 
The largest systematic errors are due to the understanding of the
detectors, especially the optics, which are under continuing study.
(Knowledge of these has been included in our published 
science results~\cite{oser00b} )

\subsubsection{Gamma-Ray Energy Threshold}

The energy threshold of a detector has to be defined before
it can be calculated.
Because most detectors, especially those measuring air showers,
do not have a sharp turn-on in sensitivity, assigning an
energy threshold is a subtle task.
Here we use the concept of the \lq spectral threshold energy\rq~
which is the energy at which the gamma-ray detection rate per 
unit energy reaches its peak. 
It is defined as the energy at which $A_{eff}(E)dN/dE$ is at a maximum.
As such it depends on $dN/dE$, the differential spectral flux of the 
gamma-ray source, and is therefore source dependent.

For this paper we assume a differential spectral flux: $dN/dE \sim E^{-2.4}$
which is similar to that measured for the Crab Nebula in the $200-1000$ GeV
region.
Convolving this spectrum with an average of the effective area curves
shown in figure~\ref{eff_area} gives 190 +/- 60 GeV as the spectral threshold
energy.
These numbers do not change significantly for spectral indices between 
2.3 and 2.5.

\subsubsection{Hadron Rejection}

Another important feature of a gamma-ray detector is its hadron 
rejection capability. 
STACEE achieves hadron rejection at the trigger level (a factor of 50-75)
essentially by an on-line topology cut given by the L1/L2 criteria.
Offline, another factor of between 2 and 3 is achieved by more precise
timing cuts. 
These figures are estimated by comparing the event rates obtained (mostly
due to hadrons) with those expected in the event of no rejection 
(calculated from known cosmic ray fluxes).

One figure of merit used to quantify a detector's capability is the 
sensitivity to the Crab. 
Using our published result on this source~\cite{oser00b} we can say 
that STACEE-32 could detect the Crab at a 5$\sigma$ significance level
in 23 hours.  

\section{Conclusions}

STACEE-32 was a first-generation heliostat Cherenkov  detector. 
It used 32 heliostats from an array of 212 heliostats to collect  
Cherenkov light from air showers generated by the impact
of high energy gamma rays on the upper atmosphere. 
The large collection area allowed operation at a lower energy threshold 
than previously obtained by ground based detectors.
This device was operated for a complete observing season (1998-99)
during which time a strong gamma-ray 
signal from the Crab Nebula was obtained \cite{oser00a}, \cite{oser00b}.

STACEE-32 served as a proof-of-concept experiment. 
We are now completing an upgrade of the detector to its design 
configuration.
Features of the upgrade include:

\begin{itemize}

\item{the heliostat field (asphalt) has been darkened to reduce background
from albedo of night sky light,}

\item{new custom trigger electronics with less dead-time allow operation at
lower thresholds,}  

\item{twice as many heliostats are used, for a greater light collection area,}

\item{a new Linux-based data acquisition system with much reduced dead-time
has been installed, and}

\item{state-of-the-art 1 GS/s FADCs have been introduced to improve
timing and energy measurements on all channels.}

\end{itemize}
 
\section{Acknowledgements}

We are grateful to the staff at the NSTTF for their excellent support.
We acknowledge the loan of electronic equipment from the Physics 
Division of Los Alamos National Laboratory.
We also thank the SNO collaboration
for providing us with acrylic for light concentrators, and the machine
shop staffs at McGill and Chicago for their expert assistance.  
Many thanks
to Tom Brennan, Katie Burns,
Jaci Conrad, Cathy Farrow, William Loh, Anthony Miceli, Gora Mohanty,
Alex Montgomery, 
Marta Lewandowska, Heather Ueunten, and Fran\c{c}ois Vincent.  This work
was supported in part by the National Science Foundation, the Natural
Sciences and Engineering Research Council, FCAR (Fonds pour la
Formation de Chercheurs et l'Aide \`a la Recherche), the Research
Corporation, and the California Space Institute.

\end{document}